\RequirePackage{fix-cm}
\documentclass[twocolumn,epjc3]{svjour3}  
\smartqed  
\RequirePackage{graphicx}

\journalname{Eur. Phys. J. A}

\usepackage[english]{babel}
\usepackage[utf8]{inputenc}
\usepackage{amsmath}
\usepackage{graphicx,color,epstopdf,grffile,upgreek}
\usepackage[bookmarks,pdfhighlight=/O,pdfstartview=FitH]{hyperref}
\usepackage{mathrsfs,amssymb,amsopn,bm,bbm,slashed,color}
\usepackage{xspace}
\usepackage[super]{nth}
\RequirePackage{multirow}

\title{QCD Challenges from \pp to \AAA Collisions}

\author{J.~Adolfsson\thanksref{addr1}, A.~Andronic\thanksref{addr2},
  C.~Bierlich\thanksref{addr3,addr17}, P.~Bozek\thanksref{addr4},
  S.~Chakraborty\thanksref{addr3}, P.~Christiansen\thanksref{e1,addr1},
  D.~D.~Chinellato\thanksref{addr5}, R.J.~Fries\thanksref{addr6}, G.~Gustafson\thanksref{addr3}, H.~van
  Hees\thanksref{addr7}, P.~M.~Jacobs\thanksref{addr8}, D.~J.~Kim\thanksref{addr9}, L.~L\"{o}nnblad\thanksref{addr3}, M.~Mace\thanksref{addr9,addr19},
  O.~Matonoha\thanksref{addr1}, A.~Mazeliauskas\thanksref{addr10,addr11}, A.~Morsch\thanksref{addr11},
  A.~Nassirpour\thanksref{addr1}, A.~Ohlson\thanksref{addr1},
 A.~Ortiz\thanksref{addr12}, A.~Oskarsson\thanksref{addr1}, I.~Otterlund\thanksref{addr1},
  G.~Pai\'{c}\thanksref{addr12}, D.V.~Perepelitsa\thanksref{addr13}, C.~Plumberg\thanksref{addr3}, R.~Preghenella\thanksref{addr14}, R.~Rapp\thanksref{addr6},
  C.~O.~Rasmussen\thanksref{addr3}, A.~Rossi\thanksref{addr15},
  O.~V.~Rueda\thanksref{addr1}, A.~V.~D.~Silva\thanksref{addr5},
  D.~Silvermyr\thanksref{addr1}, A.~Timmins\thanksref{addr16}, T.~Sj\"{o}strand\thanksref{addr3},
  R.~T\"{o}rnkvist\thanksref{addr3}, M.~Utheim\thanksref{addr3},
  V.~Vislavicius\thanksref{addr17}, U.~A.~Wiedemann\thanksref{addr11}, K.~Zapp\thanksref{addr3}, W.~Zhao\thanksref{addr18}}

\thankstext{e1}{e-mail: peter.christiansen@hep.lu.se}

\institute{Division of Particle Physics, Lund University, Lund,
Sweden \label{addr1}
\and
Institut f\"ur Kernphysik, Westf\"alische Wilhelms-Universit\"at M\"unster, M\"unster, Germany \label{addr2}
\and 
Division of Theoretical Particle Physics, Lund University, Lund, Sweden \label{addr3}
\and 
Faculty of Physics and Applied Computer Science, AGH University of Science and
Technology, Krakow, Poland \label{addr4}
\and
Universidade Estadual de Campinas (UNICAMP), Campinas, Brazil \label{addr5}
\and 
Cyclotron Institute and Department of Physics and Astronomy, Texas A\&M
University, College Station, Texas, United States. \label{addr6}
\and
Institut f\"ur Theoretische Physik, J.W.~Goethe-Universit\"at, Frankfurt am Main, Germany \label{addr7}
\and 
Lawrence Berkeley National Laboratory, Berkeley, California, United States \label{addr8}
\and 
Department of Physics, University of Jyv\"{a}skyl\"{a}, Jyv\"{a}skyl\"{a}, Finland \label{addr9}
\and
Helsinki Institute of Physics, University of Helsinki, Helsinki, Finland \label{addr19}
\and
Institut f\"ur Theoretische Physik, Ruprecht-Karls-Universit\"{a}t Heidelberg,
Heidelberg, Germany \label{addr10}
\and
European Organization for Nuclear Research (CERN), Geneva, Switzerland \label{addr11}
\and
Instituto de Ciencias Nucleares, Universidad Nacional Aut\'{o}noma de
M\'{e}xico, Mexico City, Mexico \label{addr12}
\and
University of Colorado Boulder, Colorado, United States \label{addr13}
\and
INFN, Sezione di Bologna, Bologna, Italy \label{addr14}
\and
INFN, Sezione di Padova, Padova, Italy \label{addr15}
\and
University of Houston, Houston, Texas, United States \label{addr16}
\and
Niels Bohr Institute, University of Copenhagen, Copenhagen,
Denmark \label{addr17}
\and
Peking University, Beijing, China \label{addr18}
}

\newcommand{\eg}{\emph{e.g.}}
\newcommand{\ie}{\emph{i.e.}}

\newcommand{\beq}{\begin{equation}}
\newcommand{\eeq}{\end{equation}}
\newcommand{\bea}{\begin{eqnarray}}
\newcommand{\eea}{\end{eqnarray}}

\newcommand{\gsim}{\gtrsim}

\newcommand{\pp}{pp\xspace}
\newcommand{\pA}{p--A\xspace}
\newcommand{\pPb}{p--Pb\xspace}
\newcommand{\PbPb}{Pb--Pb\xspace}
\newcommand{\AAA}{A--A\xspace}
\newcommand{\AuAu}{Au--Au\xspace}
\newcommand{\OO}{O--O\xspace}
\newcommand{\ArAr}{Ar--Ar\xspace}

\newcommand{\PHI}{\ensuremath{\phi}\xspace}
\newcommand{\OMEGA}{\ensuremath{\Omega}\xspace}
\newcommand{\XI}{\ensuremath{\Xi}\xspace}
\newcommand{\XIM}{\ensuremath{\Xi^-}\xspace}
\newcommand{\XIP}{\ensuremath{\overline{\Xi}{}^{+}}\xspace}
\newcommand{\LA}{\ensuremath{\Lambda}\xspace}
\newcommand{\AL}{\ensuremath{\bar{\Lambda}}\xspace}
\newcommand{\KOs}{\ensuremath{\rm K^0_S}\xspace}
\newcommand{\esS}{$\eta/s$ }

\newcommand{\RAA}{\ensuremath{R_{\rm AA}}\xspace}
\newcommand{\RdAu}{\ensuremath{R_{\rm dAu}}\xspace}
\newcommand{\RpPb}{\ensuremath{R_{\rm pPb}}\xspace}
\newcommand{\RpA}{\ensuremath{R_{\rm pA}}\xspace}
\newcommand{\dndeta}{\ensuremath{\text{d}N_{\rm ch}/\text{d}\eta}\xspace}
\newcommand{\pt}{\ensuremath{p_{\rm T}}\xspace}
\newcommand{\kt}{\ensuremath{k_{\rm T}}\xspace}
\newcommand{\mt}{\ensuremath{m_{\rm T}}\xspace}
\newcommand{\gevc}[1]{\ensuremath{#1 \text{\,GeV/$c$}}\xspace}

\def\c24{c_2\left\lbrace 4 \right\rbrace}
\def\e24{\epsilon_2\left\lbrace 4 \right\rbrace}
\def\v24{v_2\left\lbrace 4 \right\rbrace}

\def\hyph{-\penalty0\hskip0pt\relax}
\def\endpar{)\penalty0\hskip0pt\relax}

\newcommand{\rr}{\ensuremath{\mathrm{R}}\xspace}
\newcommand{\Npart}{\ensuremath{N_\mathrm{part}}\xspace}
\newcommand{\ET}{\ensuremath{E_{\rm T}}\xspace}
\newcommand{\pTjetch}{\ensuremath{p_\mathrm{T,jet}^\mathrm{ch}}\xspace}
\newcommand{\TT}[1]{\ensuremath{\mathrm{TT}\{{#1}\}}}
\newcommand{\Drecoil}{\ensuremath{\Delta_\mathrm{recoil}}\xspace}

\usepackage[normalem]{ulem}

\graphicspath{{./figures/}}
\date{Received: date / Accepted: date}

\begin{document}

\maketitle
\sloppy

\begin{abstract}
  This paper is a write-up of the ideas that were presented, developed
  and discussed at the third International
  Workshop on QCD Challenges from \pp to \AAA, which took place in August 2019
  in Lund, Sweden\footnote{Workshop link:
    \url{https://indico.lucas.lu.se/event/1214/}}. The goal of the workshop
  was to focus on some of the open questions in the field and try to come up
  with concrete suggestions for how to make progress on both the experimental
  and theoretical sides. The paper gives a brief introduction to each topic and
  then summarizes the primary results.
\end{abstract}

The Quark-Gluon Plasma (QGP) is a phase of QCD matter at high temperatures in
which quarks and gluons are deconfined. As the temperature of the phase
transition is in a regime where QCD is non-perturbative, providing
quantitative theoretical calculations of QGP properties in hadronic collisions
is challenging. For this reason, the characterization of these properties has
come mainly from experimental observations in ultra-relativistic collisions of
gold nuclei at RHIC and lead nuclei at the LHC. Thus, while the QGP phase has
well defined properties in, for example, lattice QCD, the QGP paradigm used in
the research related to the experiments at RHIC and LHC is based mainly on the
need to provide a unifying explanation of many physical phenomena, the most
prominent of which include jet quenching, multi-particle long-range flow
correlations, strangeness "enhancement", charmonium/bottomonium suppression,
open heavy-flavor diffusion and electromagnetic radiation. In recent years,
QGP-like observations have also appeared to manifest themselves in small
systems~\cite{Nagle:2018nvi}. In particular, significant effects are observed
with increasing multiplicity in these collision systems, such as stronger long
range $\Delta \eta$ correlations, magnitude and sign of multi-particle flow
cumulants which are consistent with hydrodynamic flow, an enhancement in
strangeness production relative to non-strange hadrons, elliptic flow of heavy
flavor hadrons, and increasing baryon-to-meson ratios at intermediate \pt. It
remains an open and important question whether these observations require a
QGP explanation, or can be described by other physical mechanisms. At the same
time, some of the typical dense QCD medium effects, \eg, jet-quenching or
heavy flavor \RAA modification, have not yet been observed in small collision
systems. Finally, the observation of QGP-like effects in small systems also
provides new input and directions for the interpretation of the phenomena in
large systems.\\

The goal of the workshop was to discuss five concrete
topics:
\begin{itemize}
\item Can we get the initial state to reveal itself?
\item In what ways are QGP-like effects in small systems related to each other?
\item Is there jet quenching in small systems, and can we measure and calculate it?
\item How does the hadronization process depend on the properties of the hadronizing system?
\item Can heavy quarks unravel common mechanisms in small and large systems?
\end{itemize}

Each participant at the workshop was assigned to a unique topic and prepared a
poster related to this topic. First, the posters were all discussed in a
plenary session. The topical posters were then discussed within the smaller
topical groups with the goal to identify open questions and concrete ideas for
making progress on each topic. The questions and ideas were then discussed
both within the topical groups, as well as in meetings between each of the topical
groups. Finally, the main ideas and discussions were summarized in a plenary
session. This write-up is based on these final presentations and has been
organized so that each section represents one topic. The goal has been to
present each topic in a way that it can be read on its own, meaning that
sometimes information is repeated.

\section{Can we get the initial state to reveal itself?}

In this section, we consider the ways in which the initial state and
subsequent stages of a hadronic collision may imprint themselves onto
final-state observables, particularly in ways which are relevant to
distinguishing competing models of the different stages of evolution in these
collisions.\\

Small collision systems (\pp, \pA) have historically been used to study
initial and final state effects in ``cold'' nuclear matter, in order to
establish a baseline for the interpretation of heavy-ion (\AAA)
results. Comparisons with this baseline have lead to the establishment of ``hot'' medium effects in \AAA collisions, such as jet quenching \cite{Chatrchyan:2011sx}, quarkonium suppression and regeneration \cite{Chatrchyan:2012np}, strangeness enhancement \cite{ALICE:2017jyt}, and collective flow \cite{ALICE:2011ab}, all of which together provide strong evidence for the production of a color deconfined medium, the QGP.

However, in recent years collective, fluid-like features strikingly similar to
those observed in heavy-ion collisions, such as long-range correlations
\cite{Dusling:2009ni,Abelev:2012ola,CMS:2012qk} and the increase of strange
particle yields with charge particle multiplicity \cite{Acharya:2019kyh}, have
been also observed in small collisions systems. The question then arises
whether QGP is also created in these small systems or, conversely, whether
some alternative mechanism could explain the observations in all systems
simultaneously.  For example, theoretical modeling of the initial state within
the Color-Glass Condensate (CGC) / saturation physics framework \cite{Gelis:2010nm} and the subsequent space-time evolution (using kinetic theory \cite{Kurkela:2018wud} or event generators \cite{Buckley:2011ms}) has suggested alternative potential descriptions of collective phenomena in these systems, which do not require the formation of a strongly coupled, deconfined plasma which evolves hydrodynamically.

For this reason, a wide variety of different approaches have become available for modeling the stages of relativistic nuclear collisions, with each approach offering a unique way of understanding the microscopic properties of these systems.  Developing ways to discriminate between these approaches is clearly of paramount importance to the task of disentangling the origins of collectivity in nuclear collisions. In the present write-up, we consider three distinct approaches which are frequently discussed in connection with and employed in the modeling of small-system collectivity, and we propose several promising avenues for discriminating between them on the basis of theory, phenomenology, and experiment.
         
\subsection{Microscopic and macroscopic approaches}
A usual approach for constructing predictions of a given final state, is to
combine an initial state and a final state calculation. The most common final
state calculation is hydrodynamics \cite{Jeon:2015dfa}, which offers an
effective description formulating the system's dynamical and space-time
evolution in terms of relativistic fluid dynamics by coarse\hyph graining over microscopic degrees of freedom. Examples include iEBE-VISHNU \cite{Shen:2014vra} and MUSIC \cite{Schenke:2010nt}. A recent alternative description is offered by string models including interactions between strings \cite{Bierlich:2017vhg}, to allow for a similar spatio-temporal evolution in a microscopic way.
The initial state can similarly be constructed using different model
assumptions. Simple assumptions in, \eg, Pythia/Angantyr \cite{Sjostrand:2006za,Bierlich:2018xfw} or HIJING \cite{Wang:1991hta} use a smooth distribution of multi-parton interactions (MPIs) in each nucleon. More elaborate frameworks, such as the Mueller dipole formalism \cite{Mueller:1993rr,Bierlich:2019wld}, calculate a spatial distribution of gluons in the individual nucleons, and thus includes more fluctuations. A related framework is that of the CGC \cite{Gelis:2010nm}. As a framework derived formally as a high-energy effective theory of QCD, it yields classical field theory equations, renormalization group equations, etc. It is extended by modeling to a finite nucleus in the IPGlasma model \cite{Schenke:2012wb}.

These approaches may be classified according to whether they are based on microscopic or macroscopic descriptions of the system's properties, as well as whether these properties are taken to originate in the initial or final stages of the collision, as shown in Tab.~\ref{tab:my_label}.

\begin{table}[h]
    \centering
    \begin{tabular}{c|c|c|}
      & Microscopic & Macroscopic \\
      \hline
      & \multirow{3}{8em}{\centering Mueller dipoles,\\CGC}  & \multirow{3}{8em}{\centering CGC+hydro,\\QCD kin. theory} \\
      Initial state  & & \\
      & & \\
      \hline
      Final state & String interactions & Hydrodynamics\\
      \hline
    \end{tabular}
    \caption{Classifying initial- and final-state models according to whether they are based on microscopic or macroscopic descriptions of system properties.}
    \label{tab:my_label}
\end{table}{}

\subsection{Discriminators}
Our goal here is to identify several opportunities for discriminating between these different frameworks on the basis of experimental, phenomenological, and theoretical evidence, or some combination thereof.  We center the discussion around observables which we consider to be especially promising in this regard.  In particular, we consider the value in more careful analyses of multiplicity distributions, flow in small systems, and intensity interferometry in small systems.

\subsubsection{Multiplicity distributions}

Fluctuations in the total event-by-event multiplicity are sensitive in a
unique way to both the multiplicity distributions themselves as well as to the
jet pedestal (also called the Underlying Event).  Moments of these
distributions may therefore provide critical insights into the microscopic
degrees of freedom at play in large and small collisions, particularly in
their ability to constrain the still poorly understood initial state in \pp
collisions and in ultra-central nuclear collisions. Both the scaled variance
$\langle M^2 \rangle / \langle M \rangle^2$ and the total charged multiplicity
per participant nucleon, defined by $2 \langle \dndeta \rangle / \langle
\Npart \rangle$ as a function of \Npart, are excellent candidates for specific
quantities for investigating initial state fluctuations.  It was observed by
ALICE \cite{Acharya:2018hhy}, that the charged multiplicity per participant
nucleon shows an `uptick' for limiting (high) values of \Npart, breaking
participant scaling. The behavior is reproduced by some, but not all,
models. It would be beneficial to allow for a more differential study of this
effect. To allow such studies, experimentalists are highly encouraged to
publish multiplicity distributions, \pt distributions, and their joint
distributions, in the highest-multiplicity collisions. This will place strong
constraints on viable theoretical models of the initial state. To further
narrow down the sources of fluctuations needed to understand breaking of
participant scaling, theorists are similarly encouraged to attempt to
reproduce these same experimental fluctuation studies using as few independent
sources of fluctuations as possible. On both sides, these analyses should be
repeated for longitudinal correlations as a function of multiplicity and \pt (cf.\ studies
of flow-plane decorrelation), which are sensitive to both the breaking of
boost invariance and the subsequent space-time evolution of the system.
    
\subsubsection{Flow in \pp}

Current hydrodynamic approaches have difficulty in reproducing the
experimentally observed negative sign of the four-particle flow cumulant
$\c24$ in \pp collisions \cite{Zhao:2017rgg}, despite the use of a fluctuating
initial state characterized by a negative eccentricity
$\e24$~\cite{Bhalerao:2006tp}. This suggests the possibility that final-state
models (\eg, hydrodynamics), which typically generate an approximate $\v24
\propto \e24$ scaling, either fail to describe \pp collisions or fail to
exhibit the usual linear-response behavior~\cite{Zhao:2020pty}. Another
possibility is that the initial state geometry of \pp collisions is so poorly
modeled by existing approaches, that adding a correct response mechanism will
fail to reproduce data. Finally, it is also a possibility that contributions
to multi-particle correlations from non-flow sources, in spite of experimental
attempts to suppress such contributions, contaminates the signal to a degree
where a model that does not correctly add such contributions, will inevitably
fail. Several further questions can as such be addressed by the theory
community. Most importantly it should be assessed whether \emph{any} model for
final state response (hydrodynamics, string shoving, etc.) can map a negative
(toy geometry) $\e24$ to a negative $\c24$ using otherwise realistic model
values for a \pp collision. Furthermore, it should be investigated if and how
CGC or interference based calculations \cite{Blok:2017pui} can generate a
negative $\c24$ in \pp collisions. In the case of CGC, one should also understand the interplay
between initial- and final-state effects by including the response from the
final state to the $\e24$ generated by a CGC treatment.  On the experimental
side, further investigations of quantities like flow fluctuations, also in \pA
\cite{Sirunyan:2019pbr}, and symmetric cumulants in \pp
\cite{Acharya:2019vdf}, aim to further constrain models. To facilitate easier
direct comparison between the often quite involved experimental observables
and theory predictions, the use of collaborative comparison tools such as
Rivet \cite{Bierlich:2019rhm} is highly encouraged.

\subsubsection{Intensity interferometry}
The Hanbury Brown--Twiss (HBT) radii, $R^2_i$ ($i=$ out,
side, long), derived from two-hadron intensity interferometry~\cite{Lisa:2005dd}, probe both the dynamical (momentum-space) and the space-time (coordinate-space) structure of particle production in nuclear collisions.  For this reason, they may also exhibit non-trivial effects of collectivity.  Several well-documented features, present in both data and theoretical analyses, are of note, including a scaling with the transverse pair momentum (\kt), the breaking of scaling with the transverse pair mass (\mt), and the \dndeta-dependence of $R^2_i$.

Both \kt-scaling and the breaking of \mt-scaling affect the behavior of the
$R^2_i$ at large \kt.  On the one hand, \kt-scaling implies that the $R^2_i$
should decrease as $1/\kt$ in the presence of strong, transverse collective
expansion \cite{Makhlin:1987gm,Csorgo:1995vf,Csorgo:1995bi}.  On the other
hand, in the case of \mt-scaling, one expects the $R^2_i$, obtained in a
suitable coordinate system using different particle species, to exhibit
identical \mt-dependence in the absence of collective flow, but to shift apart
in the presence of fluid-like expansion \cite{Heinz:1996hm}.
    
In addition, the HBT radii are naively expected to scale linearly with
$\left(\dndeta\right)^{1/3}$ at fixed pair momentum, and this behavior has
been observed experimentally in virtually all collision systems from \pp to
\AAA \cite{Aamodt:2011kd,Adam:2015pya}.  However, the slopes exhibit a strong
hierarchy ($\text{out} < \text{side, long}$) in \pp which is observed to a
limited extent in \pA and not at all in \AAA, where the slopes are comparable
in magnitude for the three radii.  This slope hierarchy, along with other
aspects of the \dndeta-dependence of the HBT radii, may reflect fundamental
differences of the evolution of \pp geometry with multiplicity from that of
\AAA, and therefore places non-trivial constraints on models of small-system
collectivity.
    
Furthermore, while both \kt-scaling and the breaking of \mt-scaling have
been explored and are readily understood within the context of hydrodynamics,
it is crucial for alternative approaches to reproduce these signals as well
\cite{Schenke:2014zha}.  In hydrodynamic approaches, the \dndeta-dependence of
the HBT radii reflect features of the initial state and subsequent space-time
evolution \cite{Aamodt:2011kd}. However, some features, such as the strong
hierarchy discussed above, remain not very well understood and constitute an
important open challenge to all leading approaches to modeling collectivity in
small systems.

\section{In what way are QGP-like effects in small systems related to each other?}
Here we discuss how observables associated with different phenomena are related,
and propose future studies that would potentially resolve some of the
ambiguities regarding their interpretation.

\subsection{Anisotropic flow and strangeness enhancement}
An enhancement of strange particle yields in heavy-ion collisions (relative to
minimum-bias (MB) \pp collisions) is typically associated with QGP
formation. This is due to the fact that the temperatures required for
deconfinement are higher than the mass of the strange quark, allowing
equilibration to occur quickly. The damping of observed anisotropic flow
coefficients $v_{\rm n}$ with increasing $n$ in heavy-ion collisions is consistent
with viscous damping in a system expanding hydrodynamically with a small value
of \esS (therefore a small mean free path), which is also expected for
(strongly coupled) deconfined matter. Both of these effects are observed in
high-multiplicity \pp and \pPb collisions at the
LHC~\cite{ALICE:2017jyt,Abelev:2012ola,Abelev:2013haa,Aad:2015gqa}.

However, it is critical to address whether these phenomena can emerge from
alternative (non-QGP) physical mechanisms, such as in the string picture where
deconfinement is not explicitly assumed~\cite{Gustafson}. Strings in this
context are phenomenological representations of the QCD field at large
distances. High-multiplicity events often produce regions of high string
density in the transverse plane. A mechanism has been proposed where
overlapping strings can combine to form ``ropes", and this leads to an
increase in the effective string tension, which enhances the production of
strange particles as a function of
multiplicity~\cite{Bierlich:2014xba}. Interactions between strings push them
to lower-density regions --- a string ``shoving" mechanism, which can convert
the initial spatial anisotropies into final momentum anisotropies in the
distribution of produced particles~\cite{Bierlich:2017vhg}. Therefore
measurements of kinematic observables like $v_{\rm n}$ might not be sufficient to
identify the nature of the QCD state in such collisions.

In order to discriminate between the QGP and string pictures, we propose a
number of studies. The first consists of measurements of two-particle
correlations between identified particles in small systems. In the PYTHIA
model, when implementing ropes, the associated correlation functions are
different compared to the case where ropes are not
implemented~\cite{Adolfsson}. The PYTHIA model (without ropes) tends to
describe these measured correlation functions poorly, so if the introduction
of ropes improves the agreement with data, it would add further validation to
this implementation of a string description. Generally, such comparisons also
address how ``local" or ``global'' the production of conserved quantum numbers
is, and how this could differ between the QGP, where conserved numbers can
diffuse in the deconfined state, and the string picture, which contains no
such dynamics (see also the discussion in Sec.~\ref{sec:corr}).  The second
study we propose involves measurements of higher-harmonic anisotropic flow
coefficients in small systems. These have not yet been measured at RHIC or the
LHC beyond the \nth{4} order ($v_4$). Viscous hydrodynamics has specific
predictions regarding how these coefficients should decrease for \esS values
associated with QGP formation, and comparing these predictions and those from
string shoving to data could help to discriminate between the two
approaches. We note here that string model predictions of higher harmonics and
more differential flow observables are currently underway, and a quantitative
prediction is necessary to determine if these observables can distinguish
between the underlying physics of the dynamics in small systems. Finally, the
same data could also be compared to $v_{\rm n}$ predictions that invoke a parton
escape mechanism~\cite{He:2015hfa}, which would challenge the hydrodynamic
interpretation.

\subsection{Radial flow and particle production mechanisms at intermediate \pt}
The higher mean \pt values in \pp and \pPb collisions compared to \PbPb
collisions at the same multiplicities, and the smaller femtoscopic radii in
\pp and \pPb compared to \PbPb, indicate a faster collective expansion in the
smaller systems. Blast-wave fits to the light-flavor hadron spectra enable a
transverse-expansion velocity ($\beta_{\rm T}$) to be
extracted~\cite{Mazeliauskas:2019ifr}. One might wonder whether the extracted
information from the light-flavor blast-wave fits could be used to ``predict"
the multiplicity evolution of heavy-flavor hadron spectra (with charm or
bottom quarks) in small systems. If so, this would further demonstrate that
heavier quarks participate in the collective motion of the medium, which is
something that is observed in heavy-ion collisions where QGP formation is
expected. Another issue is the particle production mechanism at intermediate
\pt, roughly in the range of \gevc{\text{1--10}}. In heavy-ion collisions, it
is believed that quark coalescence (during the phase transition from a QGP to
a hadron gas) may play an important role. It enhances baryon-to-meson ratios,
and leads to a splitting between baryon and meson $v_{\rm n}$ in this \pt
region. Predictions invoking quark coalescence describe the measured
identified particle $v_{2}$ values from \pPb collisions rather well in this
region~\cite{Zhao:2019ehg}. The same predictions implement contributions from
jet fragmentation, which become larger at higher \pt, and cause a turnover in
both $v_{\rm n}$ and the baryon-to-meson ratios as a function of \pt. The onset of this turnover may be sensitive to possible jet-quenching effects in small systems, and other mechanisms such as initial state correlations which also predict a turnover. The study of these signals in the context of string-based models require further improvements, notably implementation of non-parallel string interactions.

\subsection{Hadron production from hard processes}
Jet quenching in heavy-ion collisions leads to a suppression of the production of high-\pt particles relative to expectations from a linear superposition of nucleon--nucleon collisions (MB \pp collisions). Such a suppression is characterized by measurements of \RAA. No such suppression has been observed yet, given current experimental uncertainties, either in MB or in high-multiplicity \pA collisions relative to \pp collisions.  Naively, this is surprising, as the small mean-free-paths implied by the hydrodynamic description of anisotropic flow in high-multiplicity \pA collisions suggest that also high-energy partons interact significantly and lose energy, leading to jet-quenching effects. Clarifying this situation deserves novel and precise measurements to experimentally identify possible jet-quenching effects. A recent model that incorporates both heavy-flavor hydrodynamics and jet quenching attempts to simultaneously describe heavy-flavor $v_{2}$ and \RAA~\cite{Katz:2019qwv}. While this model has been used to calculate predictions for \OO and \ArAr collisions, no attempt has been made for \pPb collisions. Predictions for \pPb might shed more light on the level of suppression expected for heavy-flavor hadrons, and should also be extended to the light-flavor sector.

\subsection{Selection biases in small systems}

Selecting events with large multiplicities (relative to the average
multiplicity) in heavy-ion collisions leads to a selection of events with
smaller than average impact parameters. In small systems, it is believed such
selections have less of an influence on selecting geometry, and more of an
influence on selecting rare hard scatterings producing very large number of
particles. A study of near-side peak properties from two-particle correlations
of charged hadrons could explore the role of hard processes with respect to
these biases~\cite{Ohlson}. At higher particle \pt, this peak is expected to
be influenced by jet fragmentation. Experimental results in \pp collisions
show that both the amplitude and the width of the peak are fairly constant
with respect to the multiplicity. A complementary study using PYTHIA found
that the average \pt of the leading particle in a jet is also approximately
constant with multiplicity~\cite{Ortiz}. Another source of information is
provided by experimental studies of \pp events which have a charged particle
with $\pt > \gevc{7}$~\cite{Kim}. Previous theoretical work indicates such
events would have smaller impact parameters (compared to those without such a
selection), and therefore smaller initial-state
eccentricities~\cite{Frankfurt:2003td}. Naively, this would lead to smaller
anisotropic flow coefficients, assuming such flow is generated by a
hydrodynamic response to the initial-state geometry. The fact that no such
effect is observed experimentally calls for events with charged particles that
have a higher \pt to be explored in more depth. If the results from additional
studies remain the same, they 
could prove challenging to assumptions of smaller impact parameters or
hydrodynamic response.

\section{Is there jet quenching in small systems, and can we measure and calculate it?}
Jet quenching and collective flow are different manifestations of the
same underlying physics, the final-state reinteraction of the degrees
of freedom liberated in a high-energy collision, though at widely
differing momentum scales. Can we relate quantitatively the
rescattering effects on a ${\cal O}(100\, {\rm GeV})$ component
tested in jet quenching to rescattering effects on a
${\cal O}(1\, {\rm GeV})$ component revealed in collective flow? The
connection is complex, for instance due to the running of the coupling and the opening of inelastic channels with increasing momentum transfer, but its elucidation would provide a deep
understanding of the dynamics of hot QCD matter.

In addition to different characteristic momentum scales, jet quenching
and collective flow also have different characteristic spatial scales,
as discussed below. A valuable tool to explore the connection between
jet quenching and collective flow is therefore to vary the size of the
collision system, by varying the mass of the nuclear projectiles. A
systematic program with this approach has been undertaken at both RHIC
and the LHC, with particular emphasis on ``small'' systems in which
one or both of the projectiles is a proton or light nucleus.

Current collider measurements with \pp, \pA, and light nucleus--A
collision systems all exhibit clear phenomena suggestive of the
presence of collective flow, while jet-quenching effects in these
systems are smaller than current measurement uncertainties. In this
section we consider the ways in which QCD collective dynamics might have such a strong
\pt-scale-dependence that very large collective flow signals could occur in
conjunction with jet-quenching signatures which evade current
measurement limits.

It is the nature of hadronic collisions that an ensemble of events recorded
with a MB trigger comprises a broad spectrum of physics processes,
such as momentum transfer between incoming partons, final-state
multiplicities, and the like. Just as in \AAA\ collisions, precise
measurements of jet quenching and collective flow in small systems require
good control over such event features, where by ``control'' we mean a
well-justified and testable connection between experimental observables and
theoretical descriptions of the collision. We label as ``Event Activity'' (EA)
the experimentally accessible observables characterizing an event, such as
forward multiplicity or transverse energy, and carry out jet quenching and
collective flow measurements as a function of EA. The crucial question is then
the relationship between measured EA and theoretically accessible quantities
such as impact parameter or number of MPIs. Only by addressing this question
for small systems can we utilize theoretical jet-quenching calculations to
assess the expected magnitude of quenching effects in small systems, and their
connection to \AAA measurements.

Moreover, given that the tail is wagging the dog more easily in the
limit of small dogs with big tails, we need to understand the
correlations between high-\pt triggers and low-\pt multi-particle
production in small systems, and the extent to which these are driven
by final-state interactions. This study will profit from a model
implementation in which high-\pt and low-\pt particle production
arise from the same dynamics, so that degrees of freedom at some scale
need not be labeled ``fluid'' or ``jet'', but are just what they are:
degrees of freedom at that scale.

In the following sections we discuss theoretical considerations of
``smallness'' from the points of view of kinetic transport and its
fluid dynamic limit; a specific measurement of correlations between
high-$p_T$ trigger and soft multi-particle production in  small
systems; and application of the Angantyr model to this
measurement. While we focus here on specific examples, we aim to put
them in context with other existing or proposed approaches and to
sketch an approach towards more general understanding of small
systems.

\subsection{QCD transport theory}
\label{sect:QCDTransportTheory}

In kinetic-transport theory, high-\pt and low-\pt particle production can arise seamlessly from the same dynamics. Also, kinetic transport interpolates seamlessly between free streaming in the smallest and close-to-ideal fluid dynamics in the largest collision systems. These two features make kinetic transport of particular interest for a combined understanding of collective flow and jet quenching in small systems. 

Applications of kinetic transport to ultra-relativistic heavy-ion collisions
are as old as applications of fluid dynamics. In the early 1980’s when Bjorken
first studied boost-invariant fluid dynamics~\cite{Bjorken:1982qr}, Baym
investigated boost-invariant transport~\cite{Baym:1984np}. The 1990s saw the
development of the first partonic cascade codes. The phenomenological interest
in these codes diminished in the early 2000’s when it was realized that parton
cascades (with the $2\to 2$ and $2\to 3$ collision kernels used at the time)
could not account for the observed magnitude of elliptic flow.  Around the
same time, in the early 2000’s, AMY~\cite{Arnold:2002zm} developed in
finite-temperature QCD a systematically improvable effective kinetic theory
(EKT) that describes transport of QCD matter with typical occupancies much
smaller than $1/\alpha_s$ and for degrees of freedom with momenta larger than
the in-medium screening scale. This EKT collision kernel is a sum of an
elastic $2\to 2$ contribution and a Landau-Pomeranchuk-Migdal (LPM) $1\to 2$
term. It may be regarded as one limit of the collision kernel used in perturbative descriptions of parton energy loss, and it has been used in jet-quenching phenomenology (AMY-formalism).

A paradigm shift about the applicability of transport theory to flow phenomena occurred when it was realized that for the collision kernel of QCD effective kinetic theory with strong coupling constant of realistic size, hydrodynamization can occur as fast as in strongly coupled field theories~\cite{Kurkela:2015qoa}. Also, multi-stage dynamical descriptions of heavy-ion collisions that include transport theory, such as AMPT, have been shown to account for flow phenomena. These developments make it at least conceivable that a unified dynamical understanding of medium-effects in low-\pt and high-\pt particle production can be based on a transport theory with a QCD-based collision kernel. 

Within such a transport model approach, the question of how large jet-quenching effects are in small systems can be reformulated as a two-step procedure: i) How significant must the final state interactions be in the soft sector to account for the observed collective flow in small systems? ii) How strong must the destructive interference or the scale-dependence of individual interactions for high-$p_T$ processes be, such that jet quenching in small systems could so far evade detection?

\begin{figure}[t]
\centering
\includegraphics[width=\columnwidth]{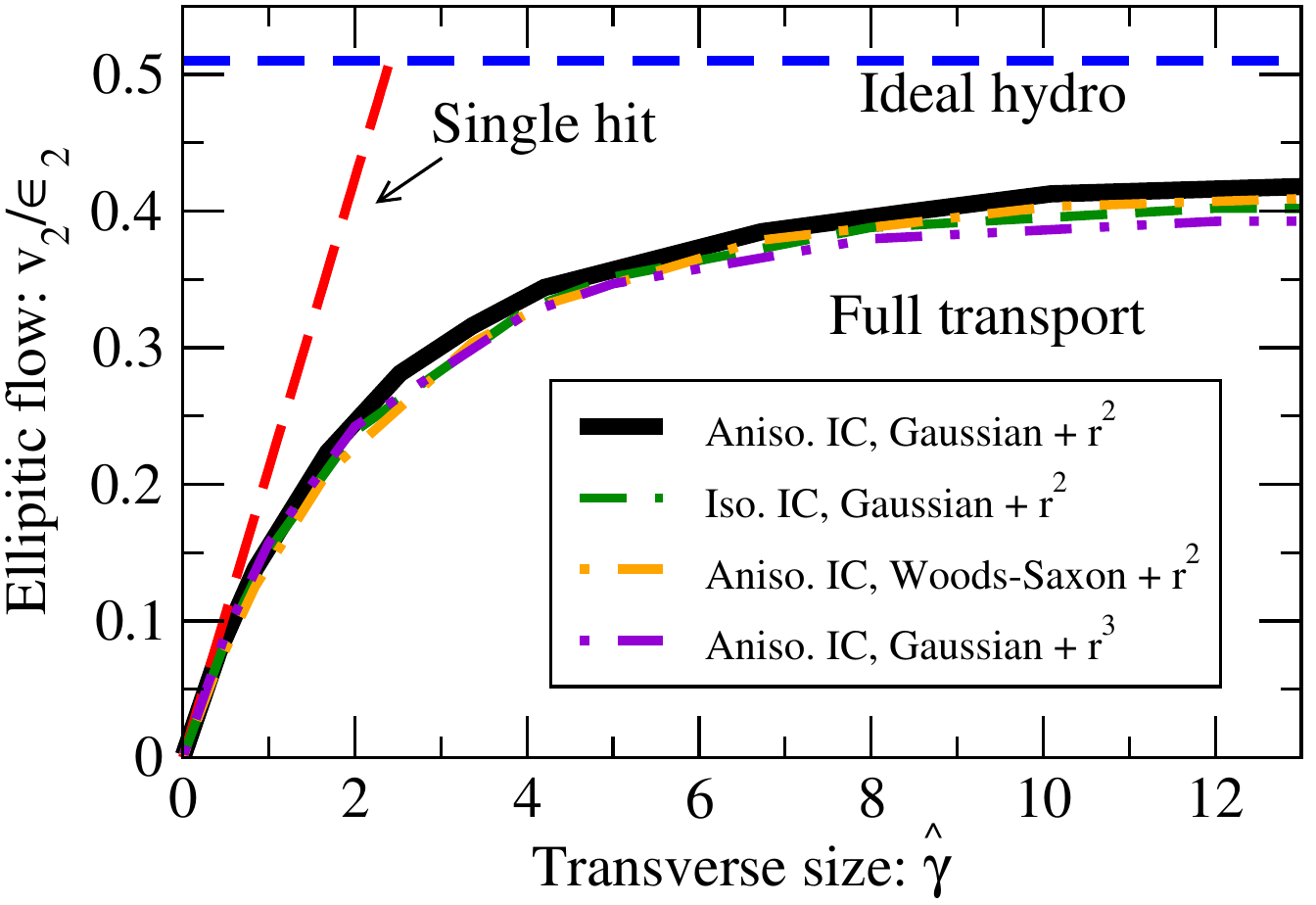}%
\caption{Linear response $v_2/\epsilon_2$ of elliptic energy flow to eccentricity as a function of opacity $\hat\gamma$ in a conformal kinetic theory. One or very few scatterings can build up significant flow in small systems. 
Figure taken from Ref.~\cite{Kurkela:2019kip}.}  
\label{Fig:FluidLimit}
\end{figure}

The first of these questions can be addressed in simplified kinetic theories
in which, \eg, the response of elliptic flow to a given eccentricity is
studied as a function of the opacity parameter, $\hat\gamma$ (a combination of
energy density and transverse extension of the system). Several studies have
established by now that even one single final-state interaction is efficient
in building up sizeable flow effects, see Fig.~\ref{Fig:FluidLimit}, and may
be sufficient to account for the observed flow in the smallest collision
systems~\cite{Borghini:2010hy,Borghini:2018xum,Kurkela:2018qeb,He:2015hfa,Sun:2019gxg}. The
question of how jet quenching fares in the limit of one or a few scatterings
has been long since identified~\cite{Tywoniuk:2014hta}, but full model studies
are still missing. The work in the present workshop did not focus on this
question. The present subsection on QCD transport theory is included here to
highlight that the open question of unifying the description of collectivity
and jet quenching across system size is currently addressed by several
approaches, including ones that differ significantly from the one documented
in the remainder of this section.

\subsection{Experimental considerations}
\label{Sect:Experiment}

\subsubsection{Jet-quenching observables}
\label{Sect:JetQuenchObs}

Jet quenching arises from the disruption of the jet shower due to
in-medium interactions. This generates several related effects for
measurements of high-\pt hadrons and reconstructed jets:

\begin{itemize}

\item ``Energy loss'': medium-induced energy transport out of the
  acceptance of the observable. For high-\pt hadrons, ``energy loss''
  corresponds to medium-induced radiation that depletes the energy of
  the (usually leading) shower branch which generates the hadron. For
  reconstructed jets it corresponds to medium-induced radiation
  transported outside the jet cone radius, \rr;

\item Medium-induced modification of jet substructure;

\item Medium-induced acoplanarity, or deflection of the jet centroid.

\end{itemize}

All three types of effect must occur if any of them does, since they
reflect different aspects of the same underlying physical
processes. It is, however, a separate set of questions which observables
are most sensitive to quenching; what aspects of quenching each are
sensitive to; and which observables offer the best control, both
experimentally and theoretically. The answers will differ depending
upon collision system and kinematic region.

Energy loss manifests itself phenomenologically as a yield suppression
relative to measurements of the same observable in an unmodified
reference system. Energy loss has been observed and measured
extensively in \AAA collisions using both inclusive and coincidence
observables, and has been studied theoretically in depth.

Study of jet quenching via substructure modification is a recent
development, based on tools adopted from the high-energy physics 
community~\cite{Andrews:2018jcm}. Initial measurements indicate a
modest medium-induced substructure
modification~\cite{Sirunyan:2017bsd,Acharya:2019djg}, though the
experimental techniques are still under development and the
theoretical interpretation of these modifications in terms of
quenching is not yet firmly established.

Medium-induced acoplanarity was the first proposed signature of jet
quenching~\cite{Appel:1985dq,Blaizot:1986ma} but has proven
challenging to observe experimentally, beyond the broadening intrinsic
to in-vacuo QCD. There are, however, new experimental and theoretical
developments in this area~\cite{Adam:2015pbpb,Adamczyk:2017yhe,Chen:2016vem}.

The observation of jet quenching in small systems would be a major
advance in our understanding of the formation and evolution of the
QGP. As noted above, the most widely explored signature
of jet quenching in \AAA\ collisions is energy loss, observed through
measurements of yield suppression with both inclusive and coincidence
observables. We now consider in turn each approach to searching for signatures of
jet quenching in small systems.

\subsubsection{Inclusive jet measurements in small systems}
\label{Sect:InclObs}

The most common observable sensitive to energy loss is \RAA, the ratio
of the yield measured in a complex collision system in which jet
quenching occurs (\eg, central \AAA\ collisions) to the scaled yield
measured in a reference system, usually MB \pp\
collisions. Scaling of the reference system yield is required to
account for the trivial geometric effect that, in the absence of
nuclear modifications, an \AAA\ collision corresponds to many
independent nucleon--nucleon collisions.

\begin{figure}[tbp]
\centering
\includegraphics[bb=286 1 570 230,clip,width=0.9\columnwidth]{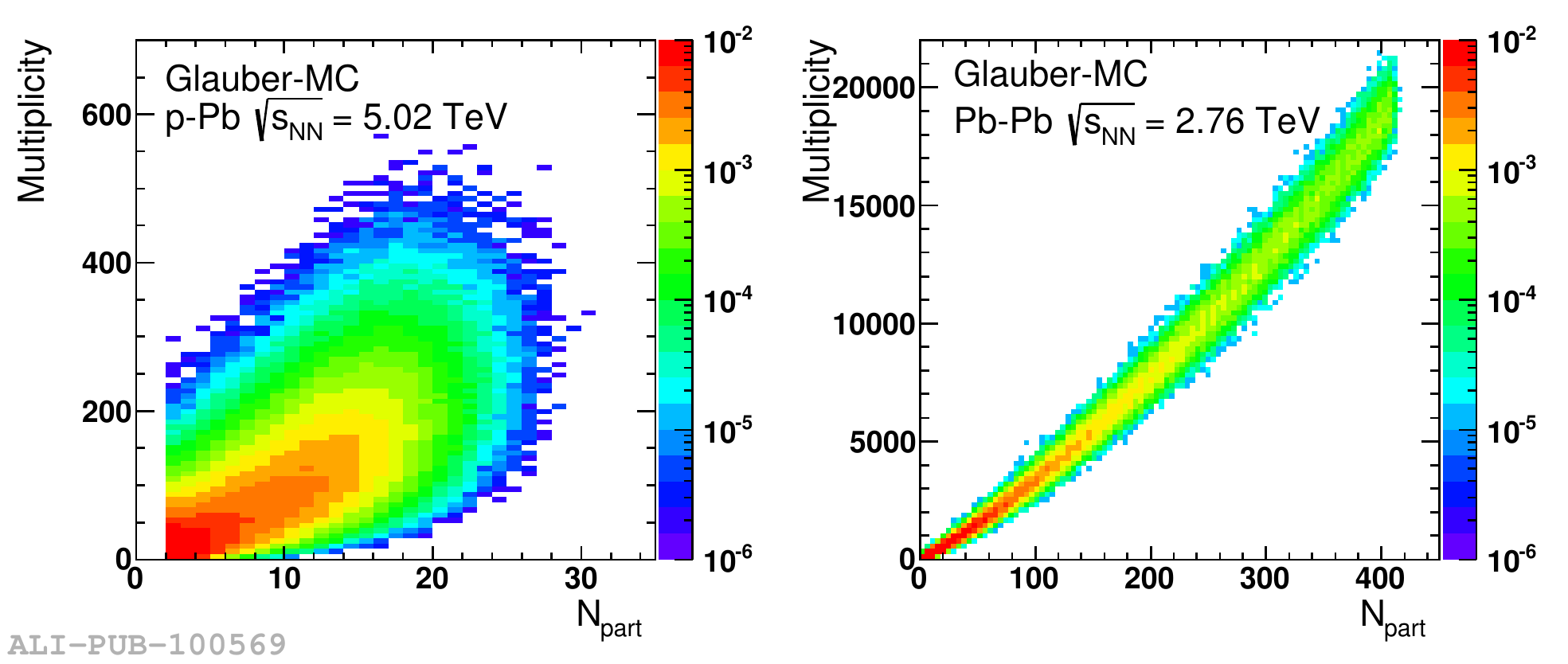}
\includegraphics[bb=1 1 285 230,clip,width=0.9\columnwidth]{2015-Sep-24-NpartMult_Glau.pdf}
\caption{Glauber-based calculations by ALICE of the correlation between forward charged-particle multiplicity (V0 detectors) and \Npart\ for \PbPb collisions (top) and \pPb collisions (bottom)~\cite{Adam:2014qja}.}  
\label{Fig:ALICEcent}
\end{figure}

In general, each \AAA\ collision can be characterized by EA (charged
multiplicity, \ET, etc.; measured at central or forward rapidities). In
Fig.~\ref{Fig:ALICEcent}, the equivalent number of nucleon--nucleon collisions
(\Npart, for ``number of participants'') is calculated using Glauber
modeling~\cite{Miller:2007ri}, which incorporates the nucleon distribution in
the nucleus and a model of EA production, and generates a correlation between
EA and \Npart. The fact that this correlation becomes blurred for decreasing
system size is a fundamental limitation for determining, on an event-by-event
basis, the transverse geometrical extent of a small collision system.

In addition, in small systems there is a well-established correlation between
the hard activity in the central region and the Underlying Event (UE), which
biases EA as an estimator of collision
geometry~\cite{Perepelitsa:2014yta,Ortiz:2018vgc}. The various collaborations
take different approaches to mitigate this effect for \pA\ collisions: ALICE
carried out model studies of the bias for different EA observables and
concluded that the forward neutron energy, measured in a zero-degree
calorimeter (ZDC), provides the least-biased EA metric~\cite{Adam:2014qja};
PHENIX applies a correction for the bias based on Monte Carlo
calculations~\cite{Adare:2013nff,Adare:2015gla}; and ATLAS does not correct
for the bias~\cite{ATLAS:2014cpa}.

\begin{figure}[tbp]
\centering
\includegraphics[width=\columnwidth]{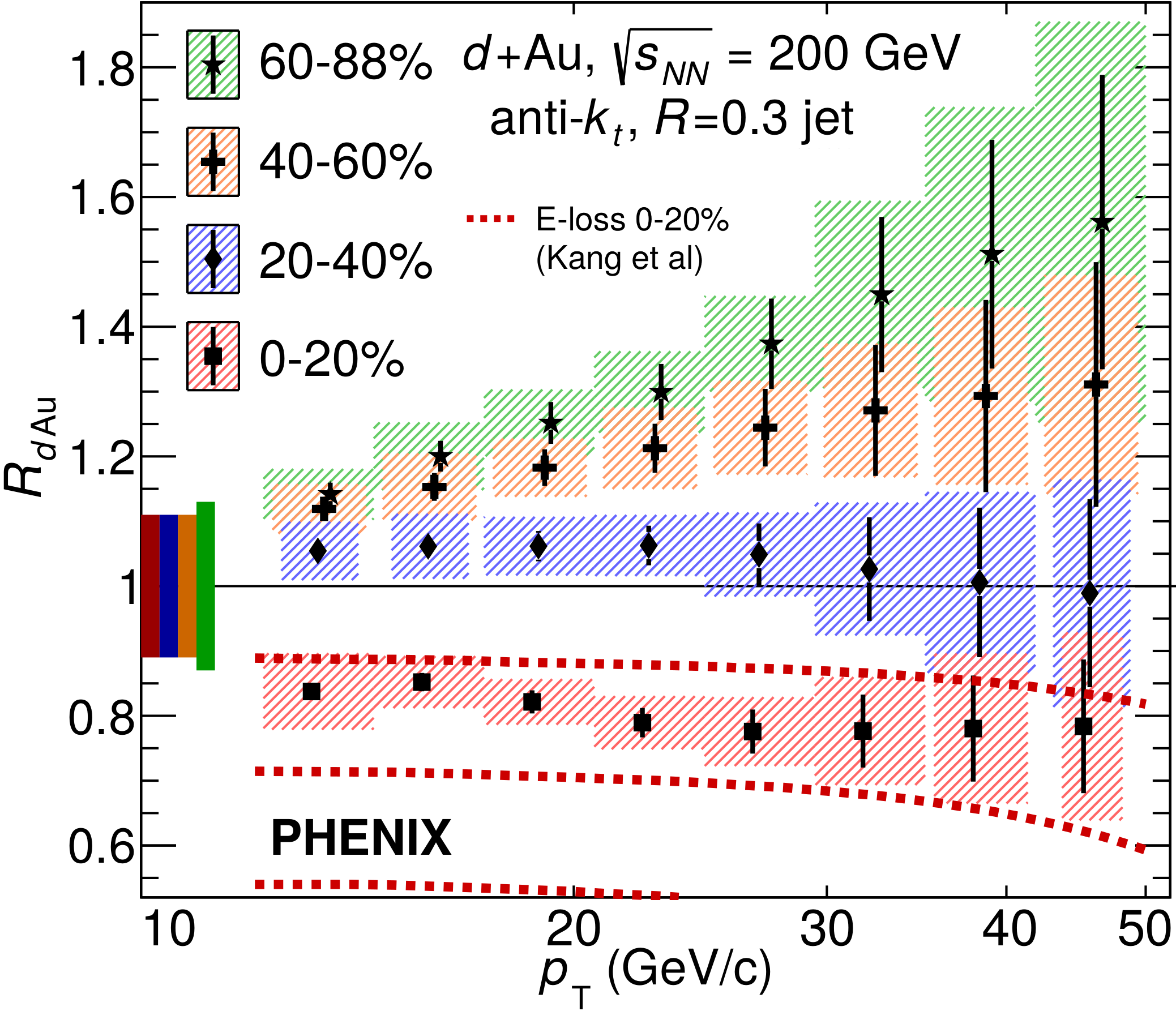} 
\includegraphics[width=\columnwidth]{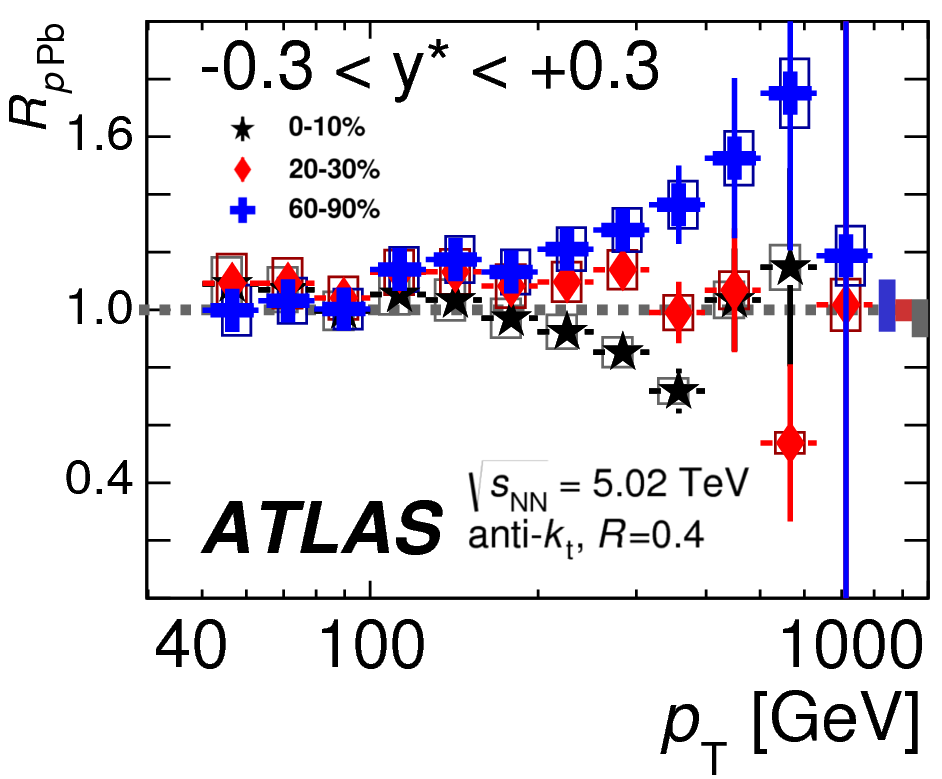} 
\caption{Inclusive jet measurements in EA-selected p/d+A collisions. Top:
  \RdAu adapted from PHENIX~\cite{Adare:2015gla}; bottom: \RpPb adapted from ATLAS~\cite{ATLAS:2014cpa}.}  
\label{Fig:PHENIXATLASRpA}
\end{figure}

Figure~\ref{Fig:PHENIXATLASRpA} shows EA-selected measurements of inclusive jet \RdAu\ by PHENIX \cite{Adare:2015gla} and \RpPb\ by ATLAS~\cite{ATLAS:2014cpa}. A strong dependence on EA is observed in both cases, with significant apparent yield enhancement for low EA and apparent yield suppression for high EA. In contrast, the MB distributions in these measurements (\ie, without EA selection) exhibit ratios \RdAu\ and \RpPb\ that are consistent with unity. Likewise, ALICE observes inclusive jet \RpPb\ consistent with unity for EA selections based on forward-neutron ZDC energy~\cite{Adam:2016jfp}. Finally, all EA-selected measurements of high-\pt inclusive hadron yields have \RpA\ consistent with 
unity~\cite{Adam:2014qja,Acharya:2018qsh,Aad:2016zif,Khachatryan:2016odn}.

It is evident that this set of EA-selected inclusive-yield measurements does not give a clear message about jet quenching in small systems. Several effects can contribute to apparent EA-dependent yield modifications, specifically initial-state effects and EA bias due to UE correlations with hard processes, in addition to jet quenching. Most importantly, EA-dependent inclusive-yield measurements have an irreducible dependence on Glauber modeling and its limitations in small systems. Further progress in the search for jet quenching in small systems requires alternative approaches. 

\subsubsection{Coincidence jet measurements in small systems}
\label{Sect:CoincObs}

We now consider a coincidence observable that is sensitive to jet quenching
but that avoids the need for Glauber modeling and the assumption that EA in
small systems is correlated with collision geometry. The observable is the
semi-inclusive distribution of jets recoiling from a high-\pt hadron, for
EA-selected \pPb\ collisions at $\sqrt{s_{\rm NN}}$ = 5.02\,TeV~\cite{Acharya:2017okq}. The EA is measured using forward charged-particle multiplicity in the Pb-going direction (V0A detector, $2.8<\eta<5.1)$. 

\begin{figure}[tbp]
\centering
\includegraphics[width=0.8\columnwidth]{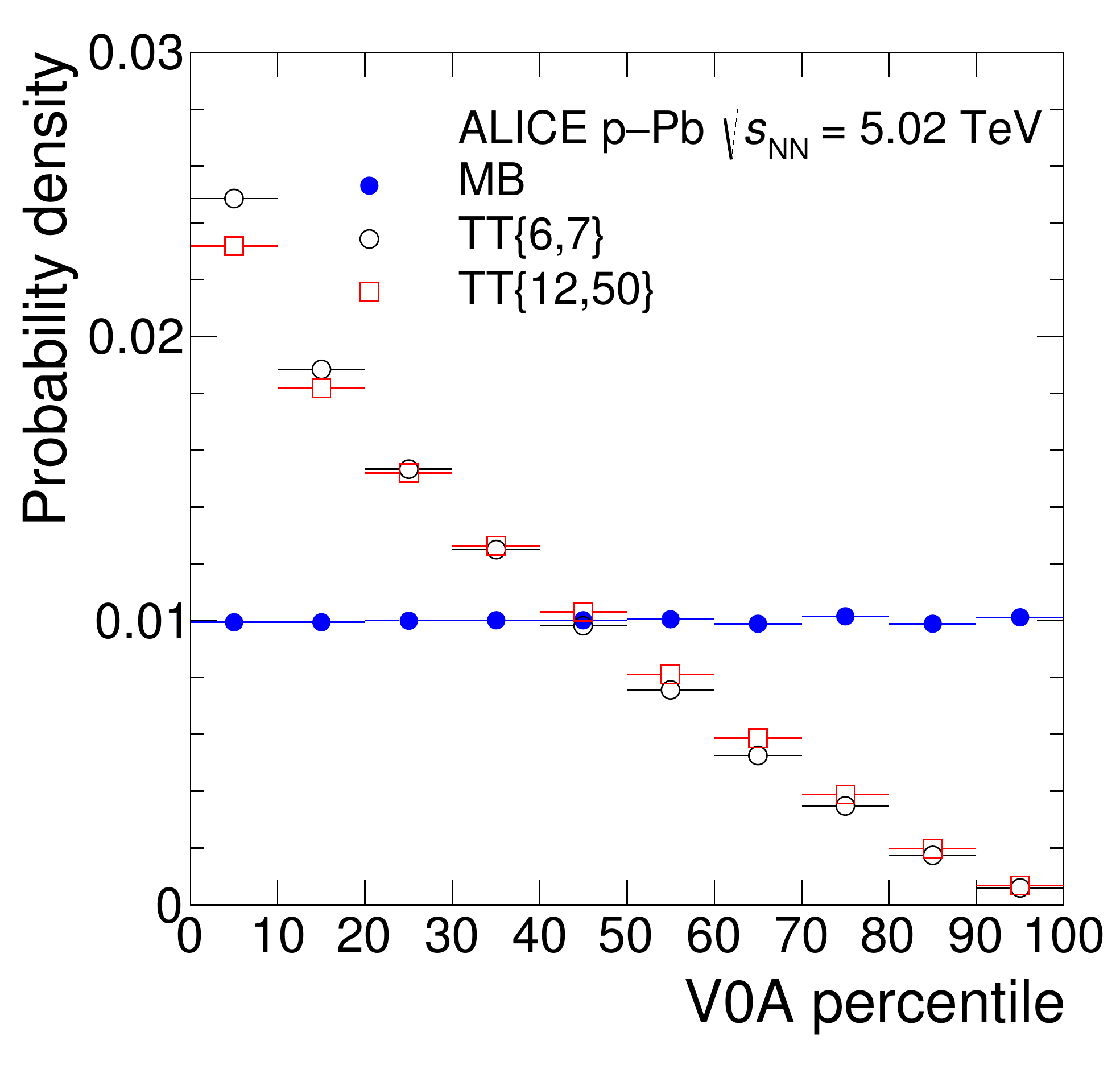}\\
\includegraphics[width=0.8\columnwidth]{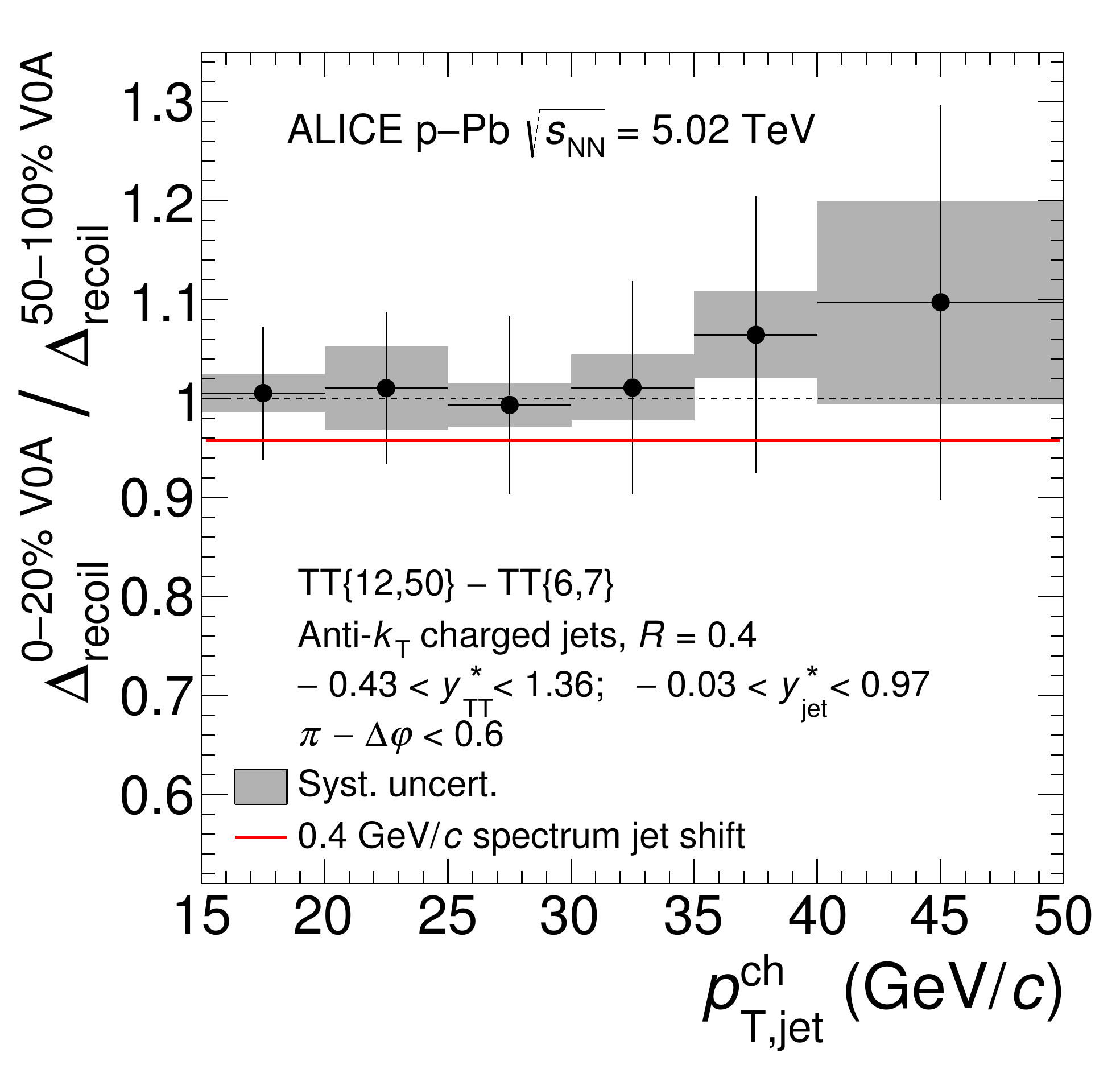} 
\caption{Figures from Ref.~\cite{Acharya:2017okq}. Top: EA(V0A) distribution for the MB population, and event populations biased by the presence of a high-\pt hadron in the central barrel with $6<\pt<\gevc{7}$ (\TT{6,7}) and $12<\pt<\gevc{50}$ (\TT{12,50}). Decile bins chosen so the MB distribution is uniform. Large V0A amplitude (large EA) corresponds to small V0A percentile. Bottom: Ratio of semi-inclusive recoil jet yield for high and low EA. Red line is the ratio if the high-EA distribution would be shifted to lower \pTjetch\ due to \pt-independent energy loss of 400 MeV.}  
\label{Fig:ALICEhJet}
\end{figure}

Figure~\ref{Fig:ALICEhJet}, top panel, shows the distribution of EA in decile
bins of the V0A distribution. The blue points show EA for the MB population,
which is trivially uniform in this representation: the decile bin boundaries
as a function of V0A amplitude are chosen precisely so that this distribution
is uniform. The red and black points show the EA (V0A) distribution for events
with a high-\pt hadron in the central barrel, which biases towards large EA. This behavior is consistent with the well-known growth of the UE for events containing jets, in this case with the UE measured in the forward direction. We discuss Monte Carlo modeling of this bias in Sec.~\ref{Sect:MCTrigBias}.

Figure~\ref{Fig:ALICEhJet}, bottom panel, shows the ratio of semi-inclusive recoil jet distributions for high-EA and low-EA event populations (see~\cite{Acharya:2017okq} for a discussion of the \Drecoil\ observable). Energy loss due to jet quenching corresponds to a shift of the recoil jet distribution to lower \pTjetch, with corresponding suppression of this ratio below unity if there are larger jet-quenching effects in high-EA than low-EA collisions. The measured ratio is consistent with unity within uncertainties, indicating no significant jet-quenching effects in this measurement. 

The measurement can, however, set a limit on jet-quenching effects, in terms
of the spectrum shift due to energy loss. Analysis, which takes into account the \pTjetch-dependence of the recoil distribution, gives a limit of 400\,MeV (90\% CL) for the population-averaged medium-induced charged-energy transport outside the jet cone radius of 0.4\,rad, for high-EA relative to low-EA events (red line in figure)~\cite{Acharya:2017okq}.

\subsection{Monte Carlo Simulations}
\label{Sect:MCSims}

In order to identify possible medium-induced jet modifications, it is
imperative to carry out thorough event generator studies to make sure that
every aspect of a measured observable is properly
understood. Especially in small systems where event activity may be
poorly correlated with the spatial extent of the medium in terms of
impact parameter or number of participants (see, \eg,
Fig.~\ref{Fig:ALICEcent}), it is important to understand possible
biases that result from the selection of events. Ideally
we would want event generators where one could switch on or off the
medium effects, and although there are several programs available for
\AAA\ (see, \eg, Refs.~\cite{Schenke:2009vr,Zapp:2013vla}), proper modeling of jets in
\pA\ and pp collisions is still lacking.

For pp collisions we have general purpose event generators such as
Herwig \cite{Bahr:2008pv}, PYTHIA
\cite{Sjostrand:2014zea,Sjostrand:2006za}, and Sherpa
\cite{Gleisberg:2008ta}, which are able to give a good description of
both hard and soft features of events. In PYTHIA\footnote{There are
  also similar models available for Herwig and Sherpa, see, \eg,
  Ref.~\cite{Bellm:2019wrh}, but we will here concentrate on PYTHIA.} this
is achieved by a MPI scenario where the
UE is treated in the same way as hard-jet production,
\ie, starting from a soft or semi-hard perturbative scattering and
adding initial- and final-state parton showers. Thus, in PYTHIA the
analogy with the dog and tail we gave in the introduction breaks down,
and we basically have a snake, where there is no clear separation into
head, body, and tail.
  
Within PYTHIA there are also (optional) models to include some collective
effects, such as \textit{color reconnections} \cite{Sjostrand:1987su},
\textit{rope hadronization} \cite{Bierlich:2014xba}, and \textit{string
  shoving} \cite{Bierlich:2017vhg}, introducing some cross-talk between
different sub-scatterings in the MPI model used to describe the UE, but none
of these specifically targets jet-quenching effects.  Nevertheless, it has
been shown \cite{ALICE:PMJ:QM19} that default PYTHIA8 (which includes color
reconnections, but no other collective effects) fairly well reproduces an
observed effect that naively would have been interpreted as enhanced jet
quenching in high-multiplicity pp jet events. Although the reason for this
unexpected result is still not fully understood, it underlines the importance
of event generators to help the interpretation of a given observable but also
highlights the care with which these results have to be interpreted. On a
similar note, recent studies of very high-multiplicity events in PYTHIA has
found jet-quenching-like features that should be further
investigated~\cite{Guy,Mishra:2019png}.

\subsubsection{The Angantyr Model}
\label{Sect:Angantyr}

For \pA\ and \AAA\ collisions there is now an event generator model implemented
in PYTHIA8 called Angantyr~\cite{Bierlich:2018xfw}. The main
ingredients are an advanced Glauber model inspired by the so-called
color fluctuation 
model~\cite{Heiselberg:1991is,Blaettel:1993ah,Alvioli:2013vk,Alvioli:2014sba,Alvioli:2014eda}
and an older generator called 
DIPSY~\cite{Avsar:2005iz,Avsar:2006jy,Flensburg:2011kk}, and the full power
of the PYTHIA8 pp machinery to describe individual nucleon--nucleon
sub-collisions. The sub-collisions are currently simply stacked
together into full \pA\ or \AAA\ events using a model, inspired by the
Fritiof program \cite{Andersson:1986gw} and the MPI model in
PYTHIA. In this way it extrapolates the dynamics of pp collisions to
heavy-ion collisions, in a fairly simple way, without involving any
collective effects between the sub-collisions. Nevertheless it is able to
describe very well the measured rapidity distribution of charged
particles in \pA\ with very few tunable parameters~\cite{Bierlich:2018xfw}, 
which then extrapolates to a very reasonable
description of multiplicity distributions in \AAA\
\cite{Bierlich:2018xfw,Acharya:2018hhy} without any further tuning.

Work has started to implement collective effects in
Angantyr, but even without these the program can still be used to
understand measurements. As an example it is shown in Ref.~\cite{Bierlich:2018xfw} 
that a non-zero $v_2$ is obtained in Angantyr,
which then can be used to understand non-collective contributions to such
an observable.

\subsubsection{Trigger bias calculations}
\label{Sect:MCTrigBias}

\begin{figure}[tbp]
\centering
\includegraphics[width=\columnwidth]{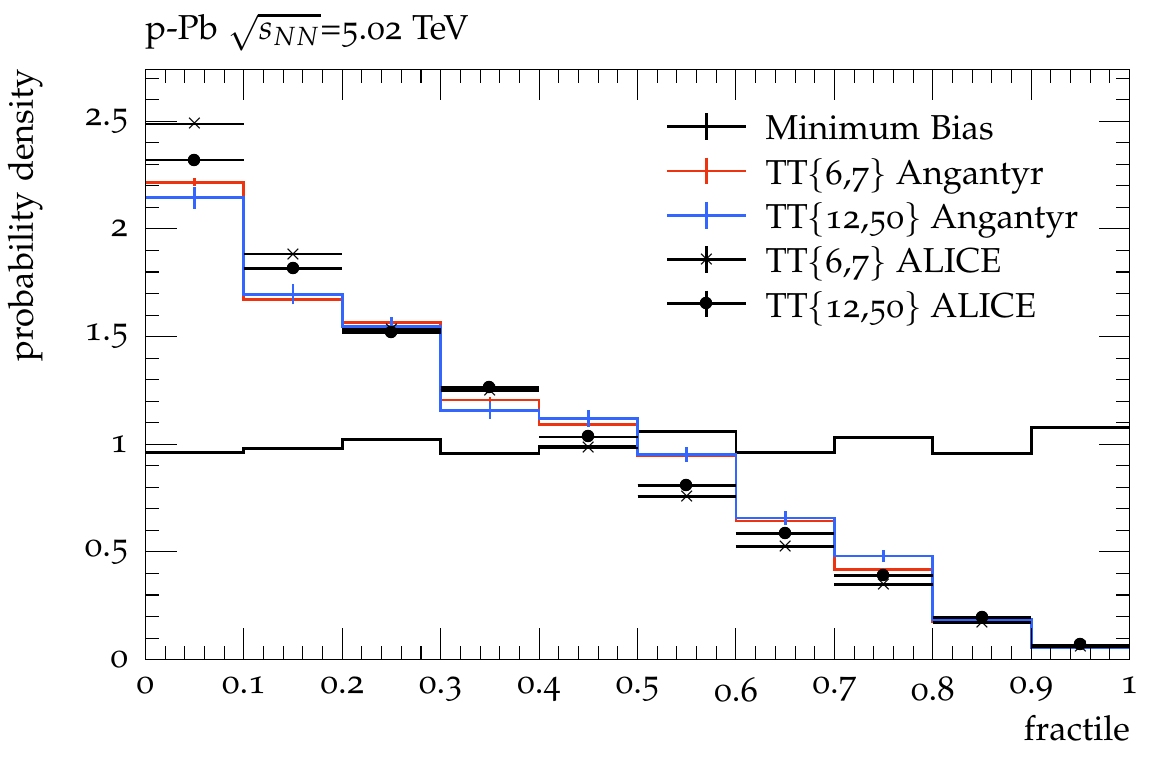} 
\caption{Trigger bias in p-Pb from Angantyr vs.~ALICE~\cite{Acharya:2017okq} (cf.~Fig.~\ref{Fig:ALICEhJet}, top panel).}  
\label{Fig:Angantyr_TrigBias_pPb}
\end{figure}

During this workshop a number of \pA\ runs were made with Angantyr to try to
shed light on the observed trigger-bias effect shown in
Fig.~\ref{Fig:ALICEhJet}. The results of the generator studies are presented
in Fig.~\ref{Fig:Angantyr_TrigBias_pPb}, and the bias found in data is very
well reproduced by the generator results.

\begin{figure}[tbp]
\centering
\includegraphics[width=\columnwidth]{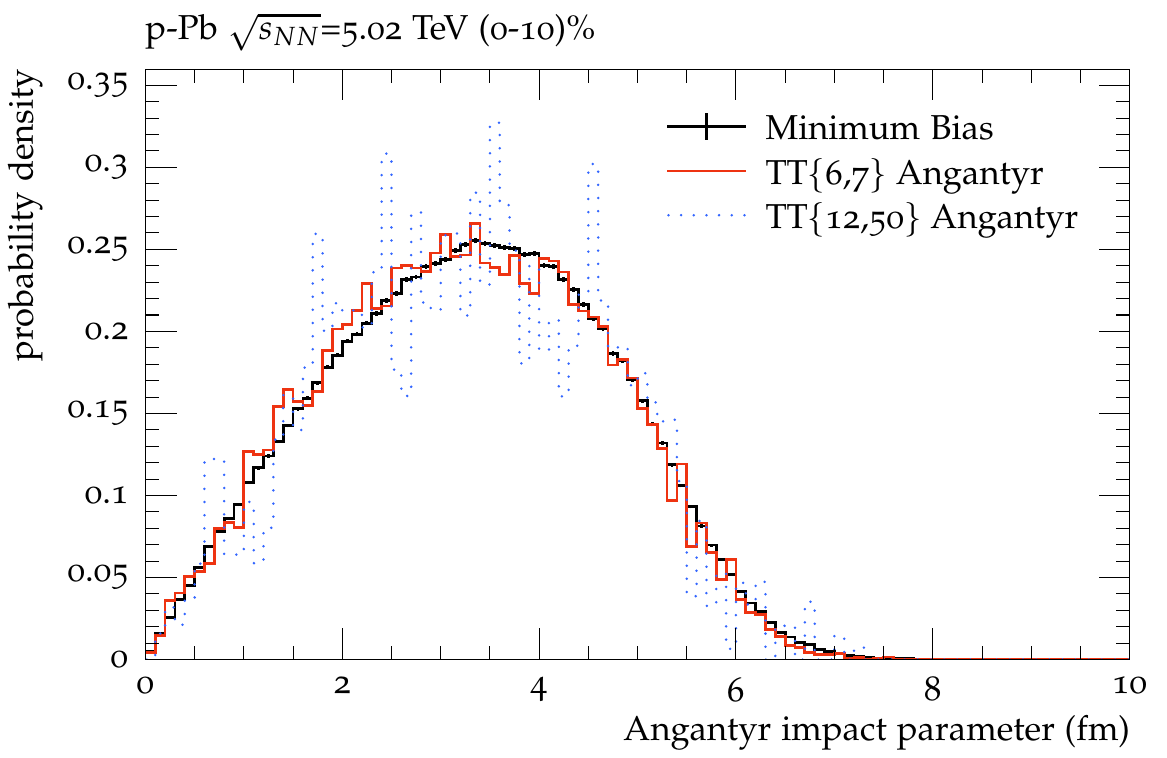}\\
\includegraphics[width=\columnwidth]{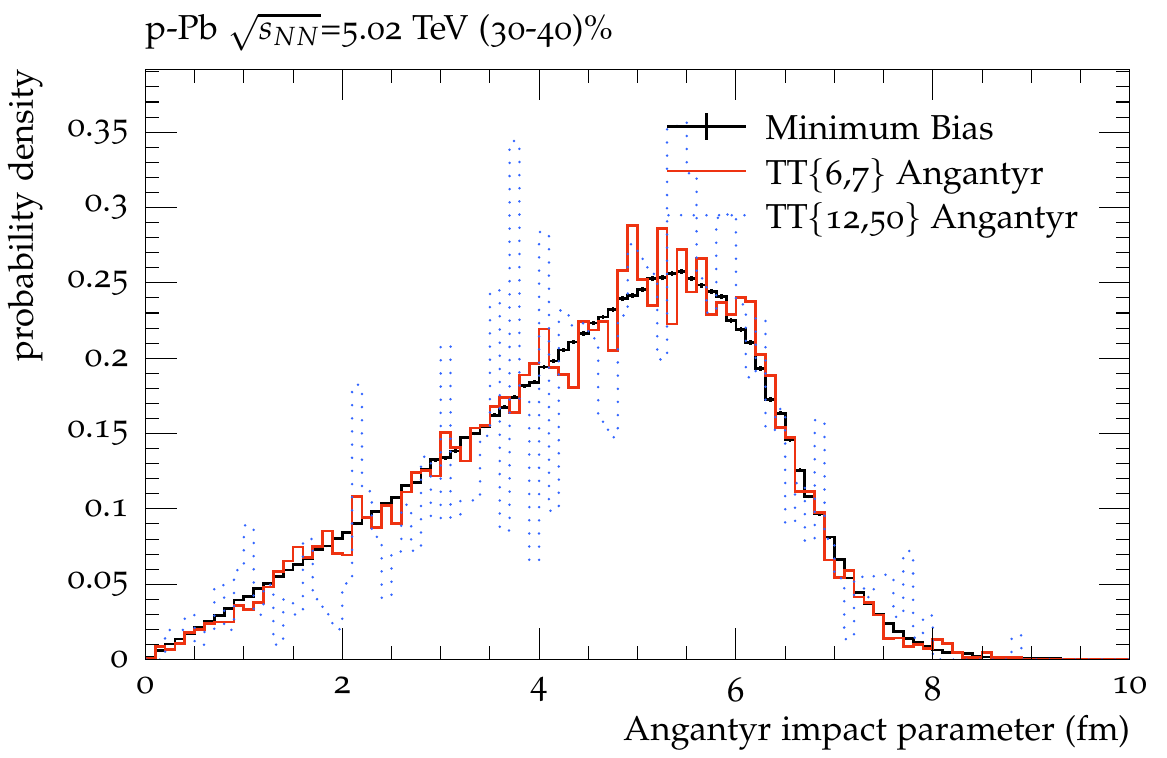}
\caption{Angantyr modeling of impact parameter distributions for two
  bins of event activity with and without a trigger hadron.}
\label{Fig:Angantyr_ImpactParam_bias_pPb}
\end{figure}

It is then possible to go into the underlying machinery of the
generator and give more support to the statement that the bias for a
\gevc{\text{12--50}} trigger hadron does not increase compared to a lower
\pt trigger of \gevc{\text{6--7}}. As an example we can inspect 
the model's impact-parameter distribution in a given
centrality decile. This is shown in
Fig.~\ref{Fig:Angantyr_ImpactParam_bias_pPb} where we first of all
see the fairly poor correlation between the centrality measure used
and the impact parameter. This comes as no surprise considering what
we already saw in Fig.~\ref{Fig:ALICEcent}, and it becomes clear
that the centrality measure is very sensitive to fluctuations in the
final state. But we also see in the figure that the impact parameter
distribution shows no significant sign of dependence on the trigger,
indicating that the bias mainly comes from final-state fluctuations in
the MPI generation rather than from the physical centrality of the
collision.

\section{How does the hadronization process depend on the properties of the hadronizing system?}
For \pp collisions, the general expectation has been that the hadronizing partons mainly reflect the initial partonic scatterings (including initial- and final-state radiation), with little or no additional final-state interactions before hadronization. Conversely, ultra-relativistic heavy-ion collisions are expected to produce a QGP in which quarks and gluons are deconfined, close to thermal equilibrium, and strongly interacting with each other up to the time of hadronization.

The observation of QGP-like effects in small systems presents two exciting opportunities: As hadronization models are challenged and new data become available, a great deal could be learned about the hadronization process itself, with profound connections to QCD confinement.
On the other hand, as our understanding of hadronization grows we might be able to peel back this layer to learn about partons and their dynamics prior to hadronization.

A key idea that we focus on in the following is that in the string picture of 
hadronization~\cite{Andersson:1983ia}, quantum numbers of produced hadrons are conserved locally. On the other hand, in a deconfined medium, or more
generally a reservoir of partons, as implied by particle production in the grand-canonical limit, quantum numbers are not expected to be conserved locally. One would then rather describe hadrons
by thermal models or by recombination of existing partons, at least some of which can come from the reservoir~\cite{Becattini:2009fv}. This idea will be discussed in the following two sections, which summarize the main outcomes of our discussions. 
A second focus point of the discussion is related to the space-time structure of jets and hadronization. This topic will be briefly elucidated in Sec.~\ref{sec:jets}.

\subsection{Identified particle yield measurements} 

\begin{figure}[!tbp]
  \centering
  \begin{minipage}[b]{0.48\textwidth}
    \includegraphics[width=\textwidth]{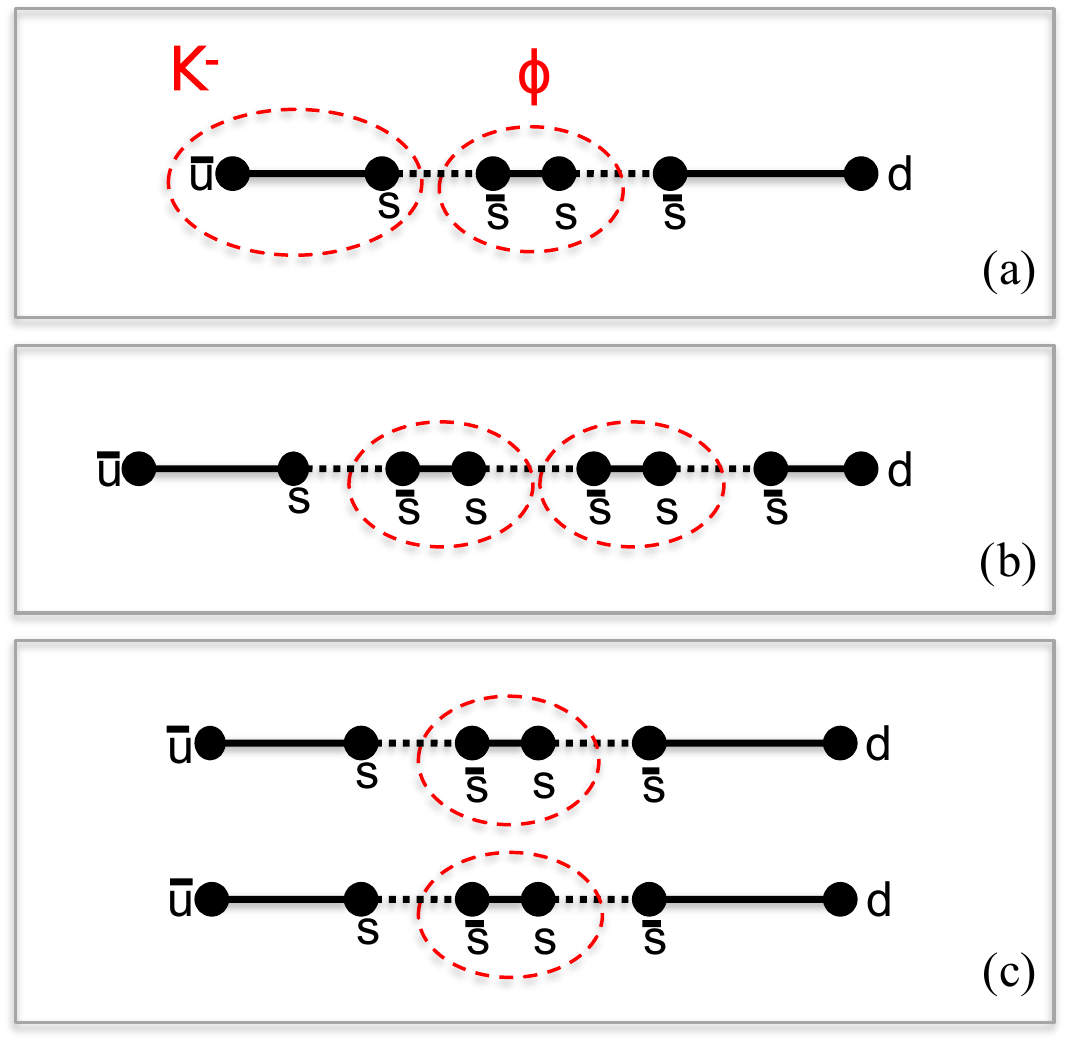}
    \caption{\label{fig:phi}Illustrations of \PHI and $\rm{K}^{-}$ strange-meson creation in the Lund string picture. (a) String breakup into a \PHI and $\rm{K}^{-}$; (b) A single string breaking up into two \PHI mesons; (c) Two strings breaking up into two \PHI mesons.}
  \end{minipage}\\
  \begin{minipage}[b]{0.48\textwidth}
    \includegraphics[width=\textwidth]{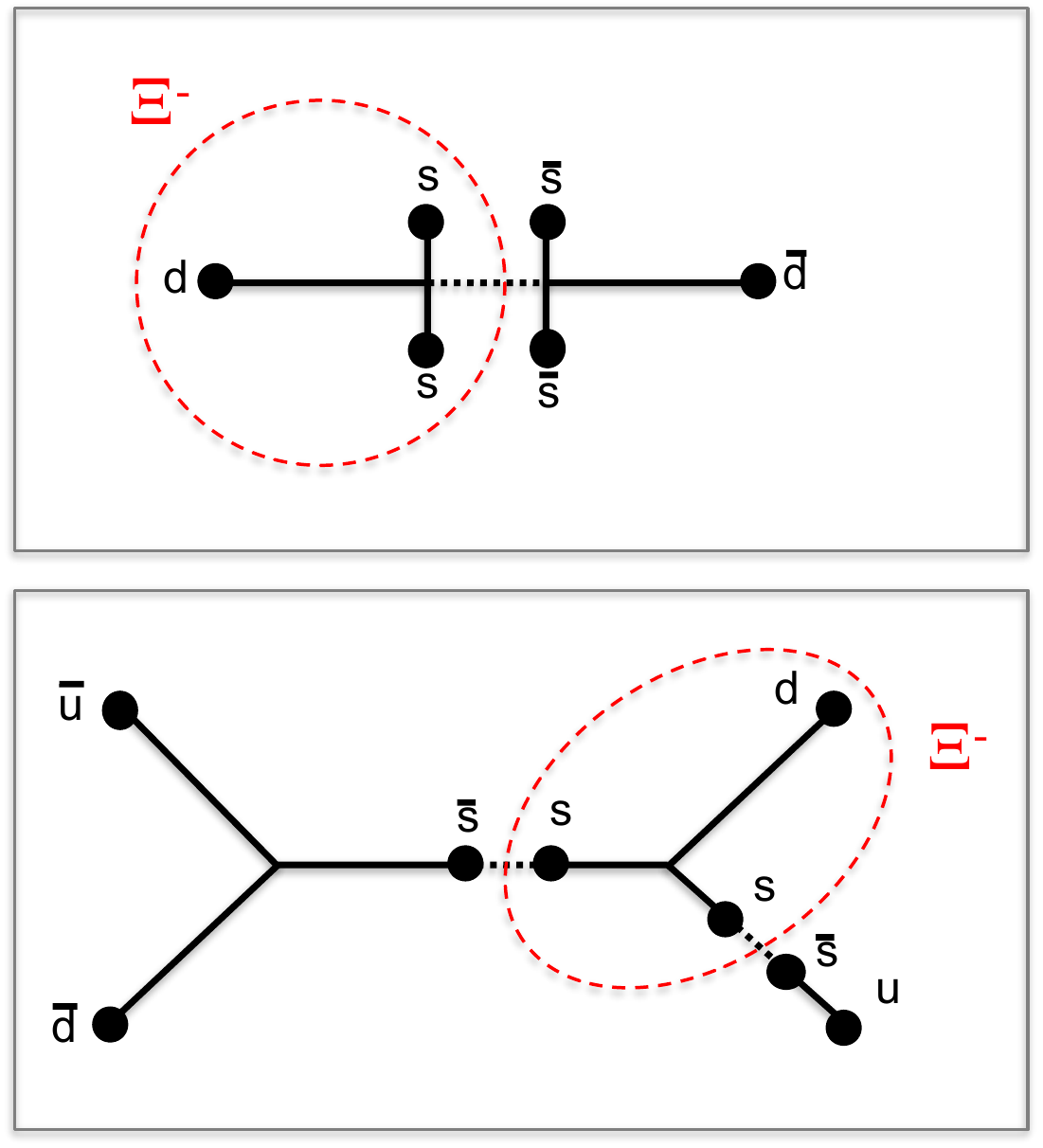}
    \caption{\label{fig:xi}Illustration of \XI production (top) in the standard Lund string framework, requiring diquark--anti-diquark pair production and (bottom) in a picture in which junctions are allowed to form.}
  \end{minipage}
\end{figure}

The \PHI meson is an interesting case when comparing string models and thermal models. In the string models, the production of a \PHI meson requires two $s\bar{s}$-string breakings and is therefore doubly suppressed. On the other hand, in basic thermal models it is treated as a non-strange particle and the production is entirely driven by its mass.

A crucial point to realize is that once a \PHI meson has been produced by a
long string, the production of a second \PHI meson next to it only requires one more $s\bar{s}$-string breaking, as illustrated in Fig.~\ref{fig:phi}. This means that for a single long string, the string model would predict the
following probabilities
\begin{multline}
P(\phi) \propto P_{s\bar{s}}^2 \quad\text{and}\quad 
P(2\phi) \propto P_{s\bar{s}}^3 \\
\rightarrow P(2\phi) \gg P(\phi)^2. 
\end{multline}
It is worth pointing out here that this kind of advanced strangeness flow in strings was part of the validation of the string model using particle production in single-diffractive events~\cite{Smith:1985fa}.

The same kind of argument is true also for multi-strange baryons. In this case, as illustrated in Fig.~\ref{fig:xi}, one must have a diquark--anti-diquark string breaking. In the case of a $ss\bar{s}\bar{s}$ breaking it is clear that one is very likely to produce \XIM and \XIP together.

For MPIs, which can lead to event-to-event variations in the number of strings, the combinatorics is more complicated but the basic idea holds and detailed calculations
can be made using event generators. One can use correlations, discussed in the next subsection, to separate correlated (same string) and uncorrelated (different string) production.

Finally, we want to add that the arguments presented here are also expected to apply for rope models~\cite{Bierlich:2014xba}. For ropes, the probability to have $s\bar{s}$ and $ss\bar{s}\bar{s}$ breakings is enhanced but the local quantum-number conservation effects are the same as for strings.

\subsection{Correlation measurements} 
\label{sec:corr}
\begin{figure}[!tbp]
  \centering
    \includegraphics[width=0.45\textwidth]{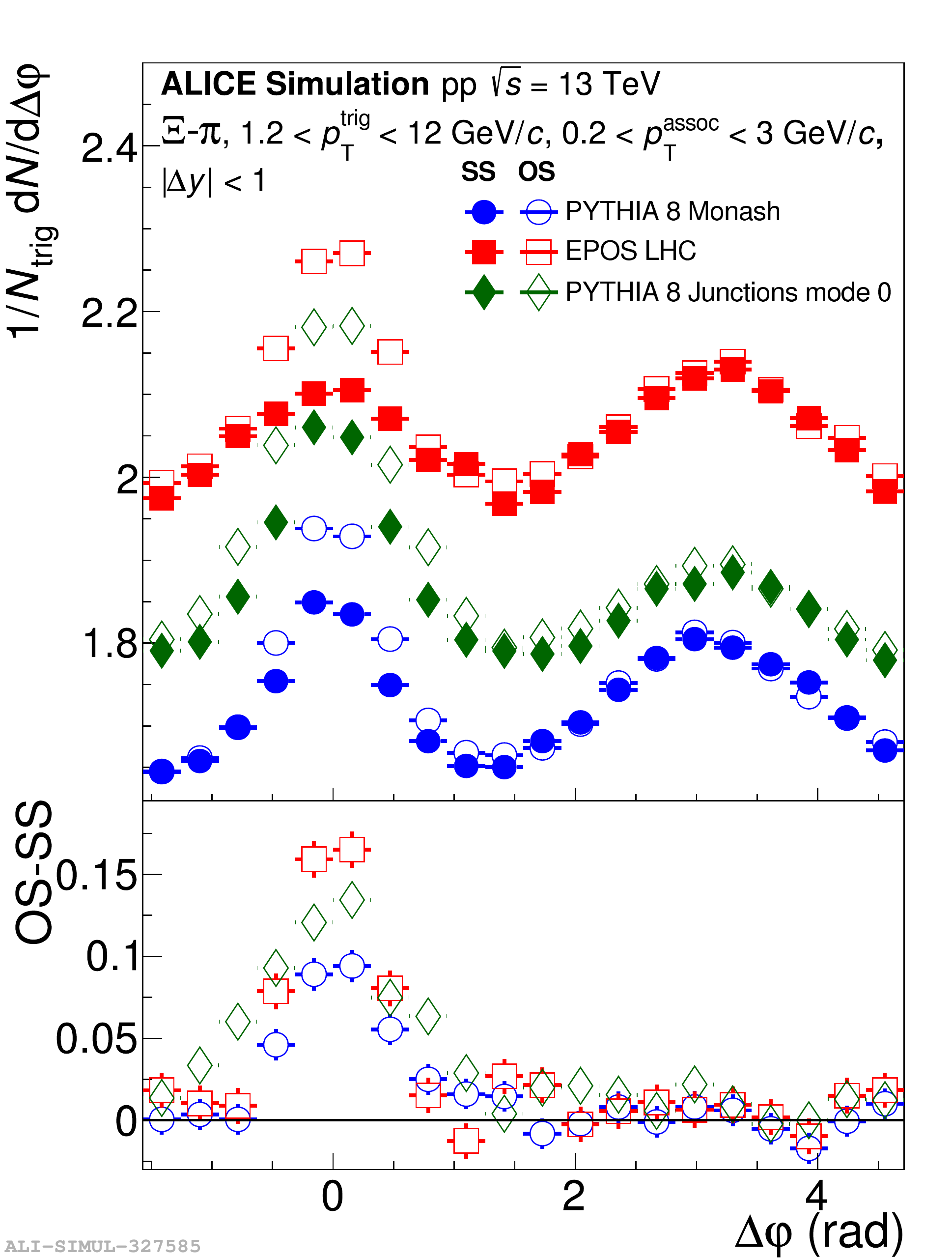}
    \includegraphics[width=0.45\textwidth]{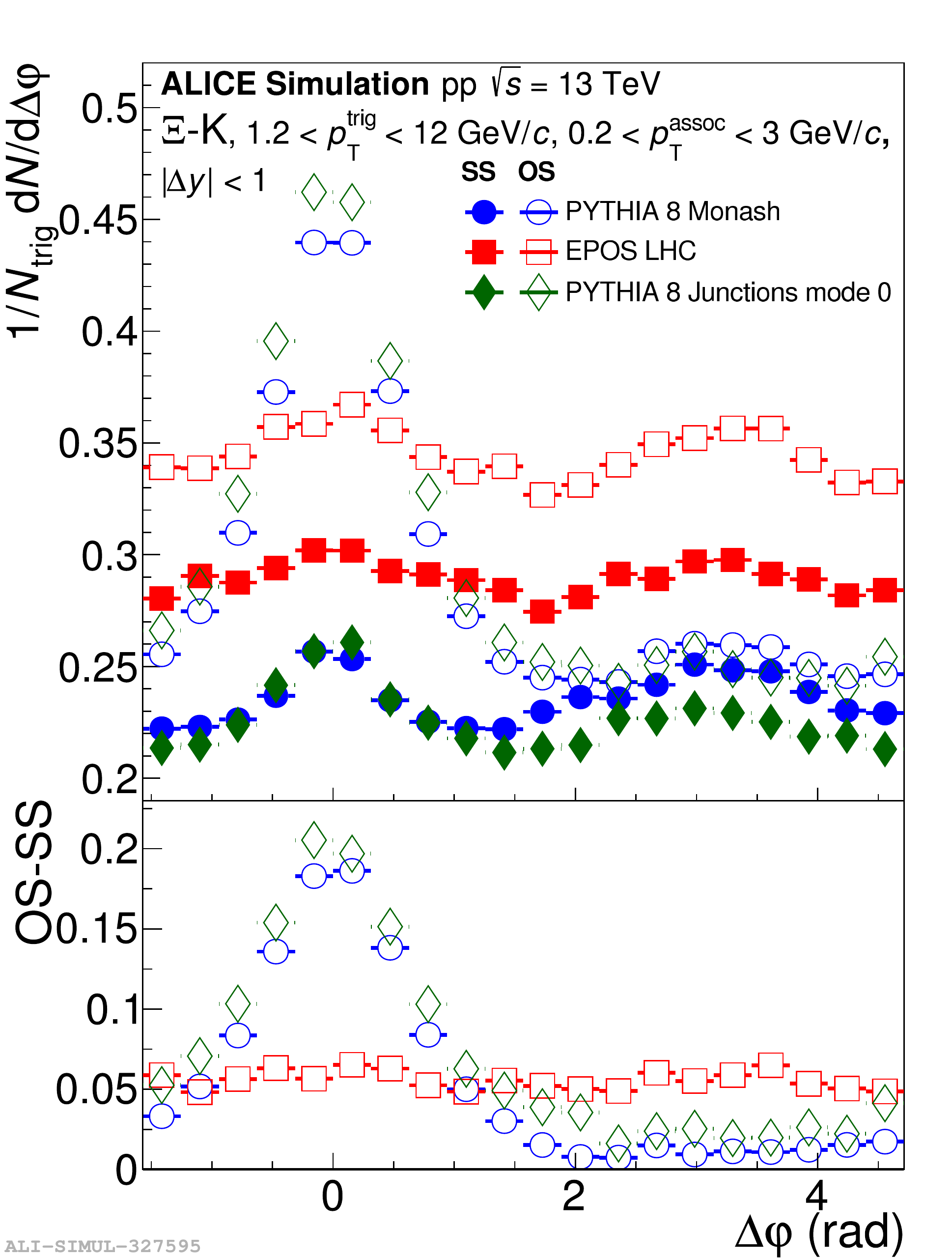}
    \caption{\label{fig:corr} Simulated $\XI$--$\pi$ correlation (top) and
      $\XI$--$K$ correlations (bottom). The upper panels show the correlations
      for opposite- (OS) and same-sign (SS) quantum numbers (charge top and
      strangeness bottom). The lower panel shows the difference (OS-SS), which
      is expected to measure the correlated production. The model
      calculations are PYTHIA8 in
      blue~\cite{Sjostrand:2006za,Sjostrand:2014zea}, EPOS-LHC in
      red~\cite{Pierog:2013ria}, and a special PYTHIA8 baryon junction tune in
      green~\cite{Christiansen:2015yqa}. Copyright CERN, reused with permission.}
\end{figure}

In this section we will refine some of the arguments of the previous
subsection. The idea is that since strangeness is conserved by the strong
interaction, one can to some degree recover the anti-strangeness associated
with strangeness production using correlation measurements\footnote{Because
  of weak decays of, for example, \KOs, strangeness is not fully conserved in
  the final state.}. Correlations of the \XIM ($ssd$) with other hadrons are
one example, explored in string fragmentation in Fig.~\ref{fig:xi}. Looking at
the top of the figure, it is clear that one needs to have at least one
$s$ quark in the diquark, so that the anti-baryon will have to be
anti-strange. This means that if string or rope models are correct, then one
should find that there are strong $\XIM$--$\AL$ correlations and
weak/vanishing $\XIM$--$\bar{p}$ correlations. One can subtract the
combinatorial correlations from these by subtracting the same-quantum-number correlations
($\XIM$--$\LA$ and $\XIM$--$p$).

In PYTHIA, an important component in the phenomenological modeling is color
reconnection, where several schemes have recently been
explored~\cite{Bierlich:2015rha}. One of the schemes that can give quite
different correlations is the baryon junction
scheme~\cite{Christiansen:2015yqa} illustrated in the bottom plot of
Fig.~\ref{fig:xi}. In the baryon junction scheme there is no longer a
"requirement" that the associated anti-baryon of a \XIM is anti-strange. In
that case, one will also produce substantially fewer pairs of \XIM and
\XIP. The important point to stress here is that there are clear
strangeness-production fingerprints in microscopic models, which can be tested
by measuring such correlations.

At the recent Quark Matter 2019 conference, ALICE has presented the first
preliminary measurements of \XI--$\pi$ and \XI--$K$ correlations. As these
results are preliminary only we show here the simulation results
(cf.~Fig.~\ref{fig:corr}; upper and lower panel, respectively). The simulation
results are shown for PYTHIA8~\cite{Sjostrand:2006za,Sjostrand:2014zea} and
EPOS-LHC~\cite{Pierog:2013ria}, which have very different production
mechanisms for multi-strange baryons, and also for the baryon junction scheme
discussed above. In PYTHIA, multistrange baryons are produced by string
breakings, while in EPOS-LHC they mainly originate from a QGP-like core. If
one focuses on the bottom panels, one observes that all three calculations
have different predictions for the pion and kaon yields associated with the
production of a \XI baryon. The simulation results demonstrate the potential
of future measurements and also highlights the care with which these
correlations will have to be modeled.

\subsection{Spacetime structure of jets} 
\label{sec:jets}
\begin{figure}[!tbp]
  \centering
    \includegraphics[width=0.45\textwidth]{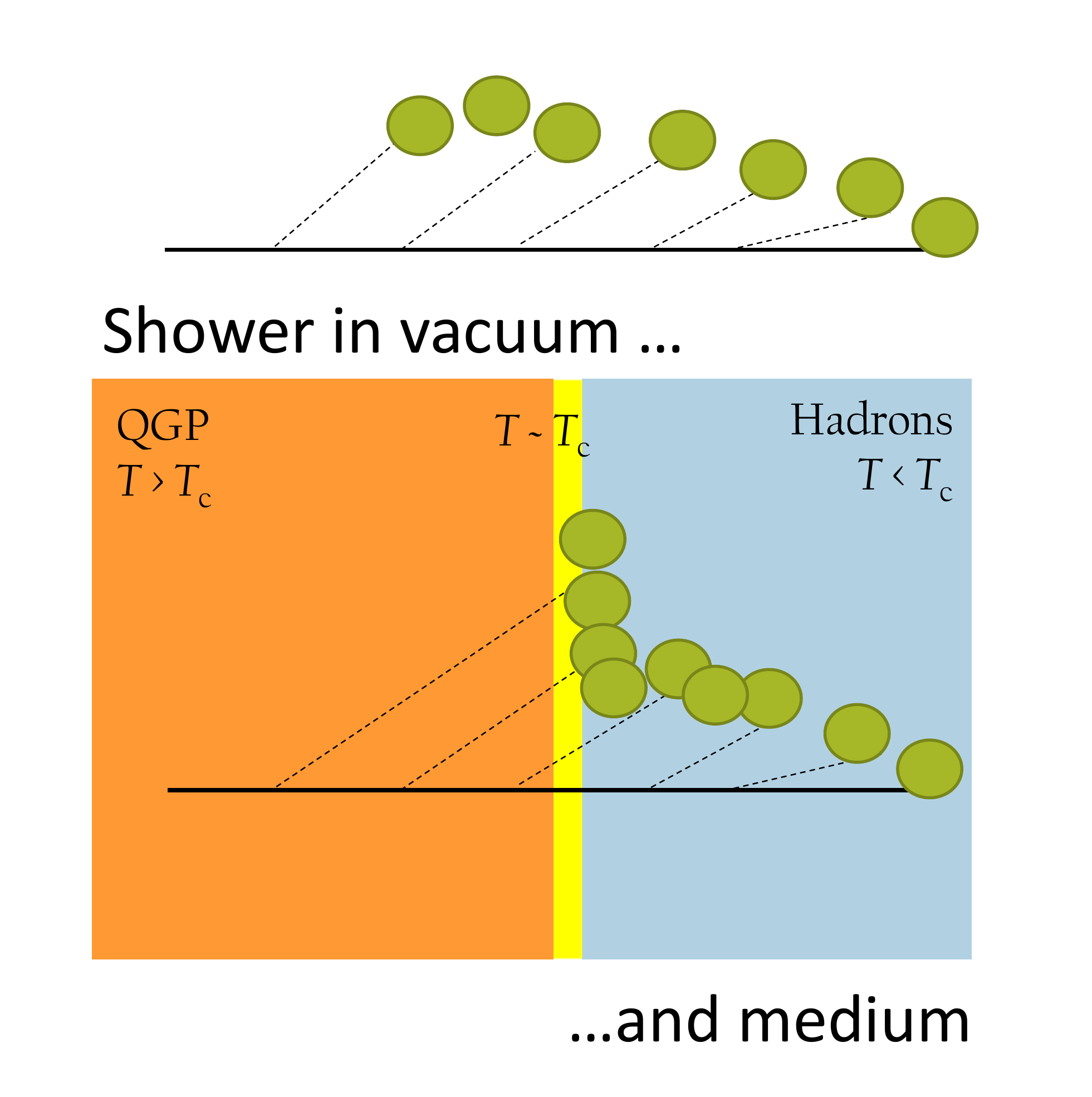}
    \caption{\label{fig:jet} Cartoon of a parton shower creating a jet in vacuum (top) and in a deconfined medium (bottom). Green dots represent partons that hadronize. In the medium, partons associated with the jet shower have to propagate to the $T=T_c$ hypersurface before hadronizing.}
\end{figure}

Jets present complementary opportunities to study QCD in general and hadronization in particular. High-momentum probes and jets have been studied over the past two decades as probes of QGP. New analyses of $e^+e^-$ data~\cite{Badea:2019vey}, \eg, using modern jet substructure techniques, can also be used to constrain hadronization models.

One novel challenge of modeling jets is the role of the space-time structure
of a jet and its relation to the space-time structure of the underlying medium
in nuclear collisions. Modern analysis techniques like re-clustering of jets
and Lund plane analyses can provide glimpses of the splitting history of jet
showers. Theoretical tools can further map these onto a space-time
picture. Hadronization in \AAA collisions is modified, as typically only part
of the jet hadronizes similar to vacuum jets, while the remaining part is
immersed in the QGP, see Fig.~\ref{fig:jet}. This is a challenge to hadronization models. Strings attached to shower partons might reconnect or terminate with thermal partons. Recombination models have been developed to model hadronization with both shower and thermal partons as input~\cite{Han:2016uhh,Fries:2019vws}.

Space-time pictures need to be implemented into improved shower Monte Carlo and hadronization models in the future. New precise jet substructure analyses in \pp, \pA,and \AAA, together with a reanalysis of $e^+e^-$ data, will be able to provide important new constraints on hadronization models. Studies of high-momentum hadrons and jets in small systems as a function of multiplicity or event
activity can be particularly illuminating.

\subsection{Summary} 
Flavor correlations between hadrons seem to provide unique fingerprints to discriminate between string and rope models, on the one hand, and thermal models and recombination, on the other. We add two additional remarks.
\begin{itemize}
    \item ALICE results suggest that strangeness production shows a very strong dependence on multiplicity~\cite{ALICE:2017jyt}. Therefore it is critical to vary the multiplicity to test if the strangeness production mechanism changes as the strangeness is enhanced (or suppression is lifted).
    \item Correlation measurements of \PHI, \XI and \OMEGA hadrons are extremely statistics hungry and therefore Run 3 and 4 at LHC are perfect opportunities for ALICE to study them.
\end{itemize}

The statistical thermal model does not \textit{a priori} give predictions for these correlations (although the notion of a correlation volume in the canonical ensemble can be introduced~\cite{Hamieh:2000tk}). The non-locality of quantum number conservation seems a natural requirement for a deconfined and thermalized QGP medium. We encourage further theoretical work into the microscopic underpinning of these models.

In the jet sector, substructure observables, as well as the dependence of observables on system size and event multiplicity, will lead to challenges to existing hadronization models. Proper modeling of space-time properties will become increasingly important.

\section{Can heavy quarks unravel common mechanisms in small and large systems?}

The heavy charm and bottom quarks ($Q=\mathrm{c}, \mathrm{b}$) play a special role in the investigation of QCD 
dynamics.  On the one hand, their masses are much larger than the typical QCD scale, $M_Q\gg \Lambda_{\rm QCD}$; 
on the other hand, their lifetimes are long enough to form hadronic bound states (although this does not hold for 
the top quark). This renders them excellent probes of: (a) hadronic structure in both open and hidden 
heavy-flavor (HF) sectors (where the large mass facilitates approximation schemes such as non-relativistic 
QCD or potential approaches); (b) particle production mechanisms in elementary collision systems (\eg, 
testing heavy-quark (HQ) production and hadronization mechanisms); (c) nuclear effects in \pA collisions 
(\eg, shadowing or absorption effects); (d) transport properties and hadronization of the 
deconfined medium in heavy-ion collisions~\cite{Prino:2016cni}.    

Several puzzling observations in the HF sector have been made in \pp,
\pA, and \AAA collisions by experiments at RHIC and the LHC in recent years that call
for investigations of seemingly related (or maybe unrelated) mechanisms.
In \pp collisions, a surprisingly large production yield of charm baryons
has been reported~\cite{Acharya:2017kfy}, with a rather significant dependence 
on rapidity~\cite{Aaij:2013mga}. A further enhancement of the
$\Lambda_{\mathrm{c}}/D^0$ ratio has been measured in \AAA collisions at
RHIC~\cite{Adam:2019hpq}, while it is less pronounced at the
LHC~\cite{alice-Lc}. For quarkonium production, an interesting dependence
on multiplicity has been measured in \pp collisions~\cite{Weber:2017hhm};
on the other hand, the enhancement in \AAA collisions was expected by
predictions of transport~\cite{Rapp:2017chc} and statistical 
hadronization~\cite{Andronic:2017pug} models, as a
consequence of (re\endpar combination of abundant anti-/charm quarks in the QGP
and/or at the hadronization transition. In \AAA
collisions, the spectra~\cite{Sirunyan:2017xss,Grosa:2018zix} and
elliptic flow~\cite{Acharya:2017qps,Sirunyan:2017plt} of HF particles
($D$ mesons and semileptonic decay leptons, and also charmonia) have
shown remarkable evidence for collectivity \textit{via} the patterns in their
nuclear modification factor (\RAA) and their large elliptic flow ($v_2$),
providing direct evidence for a strong coupling to the QGP; however, a
considerable elliptic flow for these particles has been observed as well in high-multiplicity \pPb
collisions~\cite{Sirunyan:2018toe,Acharya:2018pjd}, even though the QGP fireball,
if any, is much smaller and shorter lived; furthermore, the pertinent
\RAA shows only little modifications beyond cold-nuclear-matter
(CNM) effects~\cite{Acharya:2019mno}, except for the $\psi(2S)$~\cite{Abelev:2014zpa}.
In the following, we will report on recent discussions and insights on these issues,
specifically for the kinetics (Sec.~\ref{ssec_hf-pA}) and hadro-chemistry (Sec.~\ref{ssec_hf-hadro}) 
of open-HF particles, followed by quarkonia (Sec.~\ref{ssec_quarkonia}).

\subsection{The HF \pA Puzzle}
\label{ssec_hf-pA}
The $D$-meson $v_2$ and \RAA in 5\,TeV \pPb collisions have been
investigated in Langevin simulations for ``Brownian motion'' of charm
quarks and their hadronization assuming the presence of a
hydrodynamically expanding medium~\cite{Beraudo:2015wsd,Xu:2015iha},
with typical initial temperatures of near 400\,MeV. While the predicted
$v_2(\pt)$ can reach values close to those observed in
experiment (about 0.1)~\cite{Sirunyan:2017plt} for a sufficiently strong
$c$-quark-medium coupling, the calculated \RAA exhibits a
low-\pt peak structure (as a consequence of the collective
motion) and high-\pt suppression that does not agree with the
essentially flat dependence of the data (which in turn is consistent
with CNM effects only)~\cite{Acharya:2019mno}. One issue could be the
validity of the Langevin approximation, if only one or two rescatterings
occur in a \pA fireball. Even more extreme, it has been pointed out
that in the limit of no rescattering, an escape effect along the short
axis of the elliptic fireball can generate a positive $v_2$, although
the quantitative effect is small for heavy quarks~\cite{Li:2018leh}. 
Nevertheless, revisiting HF transport in small systems using a kinetic 
(rather than hydrodynamic) bulk medium seems to be warranted. It is also 
of interest to aim for an improved measurement of the $D$-meson $v_2$ at 
low \pt, in both \pA and \AAA collisions, which, in particular, can 
probe the presence of a negative dip as a tell-tale signature of a strong 
collective flow of heavy particles.

A recent calculation of the azimuthal asymmetry in heavy-quarkonium production in \pPb 
collisions from initial-state effects~\cite{Zhang:2019dth} -- specifically, the scattering 
of projectile partons off domains of differently oriented saturated gluon fields in the 
target nucleus -- has found an elliptic flow consistent with ALICE $J/\psi$ data. The 
$v_2(\pt)$ is predicted to be essentially identical for $J/\psi$ and $\Upsilon$. 
On the other hand, the ATLAS collaboration~\cite{Aad:2019aol} has recently measured the 
$v_2$ of semileptonic decay dimuons in 13\,TeV \pp collisions, separated into charm and 
bottom contributions and for $4 < \pt < \gevc{6}$, finding a positive signal 
for charm but values consistent with zero for bottom.
 
We also recall that in semi-central Pb-Pb collisions, the $J/\psi$ acquires a rather 
large $v_2$ of about 0.1~\cite{Acharya:2017tgv}, whereas first data for $\Upsilon(1S)$ 
$v_2$ are compatible with zero~\cite{Acharya:2019hlv,CMS:2019uhg}. These results are 
consistent with quarkonium transport models which predict a large $J/\psi$ regeneration
component at low and intermediate \pt~\cite{Rapp:2017chc}, while $\Upsilon(1S)$ 
suppression -- with a much smaller regeneration component -- happens much earlier in the 
fireball evolution where the momentum anisotropy of the fireball is still
small~\cite{Du:2017qkv,Bhaduri:2018iwr,Hong:2019ade}.  For $\pt \gsim \gevc{6}$,
however, the $J/\psi$ $v_2$ data tend to be underestimated by the transport models, 
possibly due to space-momentum correlations of fast moving anti-/charm 
quarks~\cite{He:2019vgs} which have not been included in pertinent calculations yet.

\subsection{HF Hadrochemistry}
\label{ssec_hf-hadro}
The (soft) color neutralization of (hard-produced) heavy quarks provides
an excellent window on hadronization me\-chanisms via the chemical
composition of the produced HF hadrons. For non-strange $D$-mesons ($D$ and
$D^*$), the hadrochemistry does not show significant variations
from \pp to \pPb and \PbPb collisions, and essentially follows that of a
statistical hadronization with relative weights given by thermal factors
at a ``hadronization temperature'' of $T_H\simeq160$\,MeV. The
$D_{\text{s}}/D$ ratio is also compatible with this
pattern, once a strangeness suppression factor of
$\gamma_{\text{s}} \simeq0.6$, as independently inferred from
strange-particle production, is accounted for in \pp
collisions~\cite{He:2019tik}. In semi-/central \AAA collisions
$\gamma_{\text{s}}$ is close to 1 (which is believed to be a consequence of
strangeness equilibration in the QCD fireball), and the expected
increase of the $D_{\text{s}}/D$ ratio in \AAA
collisions~\cite{Andronic:2003zv,Kuznetsova:2006bh,He:2012df} is indeed
compatible with experimental observations~\cite{Acharya:2018hre}. The
three-quark nature of baryons renders their spectrum of possible quantum
numbers substantially richer, which in the hadron spectroscopy context
gave rise to the problem of missing states in the measured spectra. Even
rather recently, it has been argued \cite{Alba:2017mqu} that QCD
thermodynamics as computed in lattice QCD (lQCD) requires more
strange-baryon states than currently listed by the particle data group
(PDG)~\cite{Tanabashi:2018oca}.

The knowledge of excited baryon states is much scarcer in the charm
(and bottom) sector. Recent measurements by ALICE in 5 and 7\,TeV \pp
collisions at midrapidity have found a much larger
$\Lambda_{\text{c}}/D^0$ ratio of $\sim$0.54$\pm$0.16 than
previously measured, \eg, in $e^+e^-$ annihilation
($\sim$0.1). It is also significantly larger than expectations from
string-based event generators~\cite{Bierlich:2015rha}, as well as
statistical hadronization ($\sim$0.22)~\cite{Andronic:2003zv} when 
utilizing the known charm-baryon states as listed by the PDG. In
Ref.~\cite{He:2019tik} it was shown that upon including a largely
augmented charm-baryon spectrum as predicted by relativistic quark models (also
consistent with lQCD~\cite{Padmanath:2014bxa}), the ALICE measurement
could be accounted for. Improvements in the string fragmentation scenario, by
accounting for color correlations beyond the leading-color approximation to
create several MPI sub-systems, can also facilitate 
the formation of charm baryons in \pp collisions~\cite{Christiansen:2015yqa}, 
yielding a $\Lambda_c/D^0$ ratio close to that measured at midrapidity.
A potential caveat in both descriptions is that the LHCb measurement at
forward rapidity finds a significantly smaller result, of about
0.25$\pm$0.05 in \pp~\cite{Aaij:2013mga} and
0.35$\pm$0.05~\cite{Aaij:2018iyy} in \pPb collisions.  These results
could point at a multiplicity dependence of this ratio, saturating at the
value given by the statistical hadronization model. A more microscopic understanding of how the multiplicity
affects this ratio is clearly in order. Of high interest are
measurements of additional charm baryons, which would also be very
helpful to quantify the feed-down contributions to ground states. For
example, recent data for $\Xi_c^0$ production in 7\,TeV \pp collisions
suggest a large enhancement relative to predictions of baseline event 
generators~\cite{Acharya:2017lwf}.

Let us finally comment on the current situation for charm-baryon
production in \AAA collisions. Recent STAR data~\cite{Adam:2019hpq}
suggest a further increase of the $\Lambda_{\text{c}}/D^0$ ratio in central
0.2\,TeV \AuAu collisions to $\sim$1.05$\pm$0.25, compared to
$\sim$0.5$\pm$0.1 in peripheral collisions. On the other hand, in \PbPb
collisions at the LHC, the centrality dependence is much less
pronounced, with a rather small increase (if any) from \pp to central
\PbPb collisions~\cite{alice-Lc}. It remains to be seen whether there is
a tension between these data, as theoretical predictions generally do not
expect large differences between RHIC and the LHC~\cite{He:2019vgs,Zhao:2018jlw,Plumari:2017ntm}. 
The main uncertainty in these measurements is their reach to low momenta, which is currently
limited to $\pt \gsim \gevc{2}$. An accurate inclusive yield measurement will
be pivotal for addressing medium modifications of this ratio, \eg, to
scrutinize the relation between the production mechanisms in \pp and \pPb
collisions and to better understand the redistribution of the
charm-hadron yields in \pt at hadronization; after all, the
charm-quark spectra at the hadronization transition share the same
modification due to their prior diffusion through the collectively
expanding fireball medium.  For
example, space-momentum correlations between fast-moving heavy quarks
and thermal partons in the outer parts of the expanding fireball have recently
been found to have a significant impact on both \RAA and $v_2$ observables 
of charm hadrons~\cite{He:2019vgs}.

\subsection{Quarkonia}
\label{ssec_quarkonia}

The discovery of quarkonia in the 1970's established the basic QCD force in
vacuum, with a perturbative color-Coulomb part at small distances and a linear
``confining'' potential taking over for distances of $r\gsim 0.2$\,fm. The
linear potential dominates the binding for all charmonia and bottomonia,
except the $\Upsilon(1S)$ ground state. Consequently, their production
systematics in \AAA collisions plays a key role in deducing the screening of
the confining force in the medium and, more generally, its role in the
properties of the QGP and its de/confinement transition. However, the
originally envisaged suppression signature has significantly evolved over the
last $\sim$15 years, and it became clear that (re)generation processes in the
hot QCD medium play a decisive role. The observed enhancement of $J/\psi$
production in \PbPb collisions at the LHC, relative to the large suppression
in \AuAu collisions at RHIC, was by and large predicted by both transport
models and the statistical hadronization model and corroborated by its
concentration at low \pt, as well as a large elliptic
flow~\cite{Rapp:2017chc}. Significant uncertainties in the regeneration
component remain due to current uncertainties in the total charm production
cross section in \PbPb collisions. To resolve those, low-\pt measurements of
both $D$-meson and $\Lambda_{\text{c}}$ production are essential, as also
mentioned in the previous subsection. The key quarkonium transport parameter
is the inelastic reaction rate, $\Gamma_{\cal Q}$, which is responsible for
both their suppression and regeneration. In a recent work~\cite{Du:2019tjf},
it has been shown that bottomonia are a more promising observable than
charmonia to infer the in-medium QCD force from their suppression pattern, due
to a smaller ``contamination'' from regeneration. A statistical analysis of
all available $\Upsilon$ $\RAA$ data from \AAA collisions at RHIC and the LHC
deduced a ``strong'' in-medium potential, with remnants of the confining force
surviving well above $T_{\text{c}}$. Interestingly, the same potential yields
a heavy-quark diffusion coefficient which is in the right range to account for
open-HF phenomenology~\cite{Liu:2018syc}. Indeed, the quantitative coupling of
open- and hidden-HF transport remains a challenging
problem~\cite{Gossiaux:2016htk,Greiner:2018fis}. However, it has been argued
that the $\Upsilon(2S)/\Upsilon(1S)$ ratio for central \PbPb collisions may
also be compatible with the statistical hadronization
model~\cite{Andronic:2017pug}. Detailed measurements of weakly bound states,
like $\psi(2S)$, $\Upsilon(3S)$, and even
$X(3872)$~\cite{Cho:2017dcy,CMS-PAS-HIN-19-005}, including their \pt
dependence and $v_2$, are promising observables to better disentangle the
production mechanism.

A full understanding of the production mechanism of quarkonia in \pp and \pPb
collisions remains elusive thus far. This includes the multiplicity dependence
of $J/\psi$ production, which increases substantially with $N_{\rm ch}$,
possibly stronger than linear (see the ALICE preliminary results in
Ref.~\cite{Weber:2017hhm} and references therein). The suppression of $J/\psi$
in \pPb is largely described in shadowing/saturation models,
but the larger suppression of $\psi(2S)$~\cite{Abelev:2014zpa}, which is more prominent 
for higher-multiplicity events~\cite{Adam:2016ohd}, remains not well understood to date, 
as is the case also for the observed $v_2$ of $J/\psi$ in high-multiplicity events~\cite{Acharya:2018pjd}.


\begin{acknowledgements}
  The organizers would like to thank all the participants for making a both
  inspiring and enjoyable workshop, where many interesting new ideas were
  discussed and developed, some of which have been summarized in this
  document. \\ The workshop was organized using funding for the project
  ``CLASH: Pinning down the origin of collective effects in small collision
  systems''. We want to thank the Knut and Alice Wallenberg Foundation for
  funding this project.\\ This work was further supported by the Swedish VR
  contract No.\ 2017-003 (C.~Bierlich), NCN under grant
  No.\ 2018/29/B/ST2/00244 (P.~Bozek), EU H2020 MSCA ITN grant agreement
  MCnetITN3/722104 (S.~Chakraborty and M.~Utheim), FAPESP under grant
  17/05685-2 (D.~D.~Chinellato), US NSF awards PHY-1516590 and PHY-1812431
  (R.J.~Fries), PAPIIT-UNAM under Project No.\ IN102118 (A.~Ortiz), US-DOE
  under award No.\ DE-SC0018117 (D.V.~Perepelitsa), US-NSF under grant
  no.-PHY-1913286 (R.~Rapp), and ERC H2020 Advanced Grant agreement
  No.\ 668679 (C.~O.~Rasmussen and T.~Sj\"{o}strand).
\end{acknowledgements}

\bibliographystyle{utphys}
\bibliography{bibliography}

\end{document}